\documentclass{amsart}
\usepackage{amsthm}
\usepackage{amsmath}
\usepackage{amsfonts}
\usepackage{amssymb}
\usepackage{graphicx}
\usepackage{float}
\usepackage{color}
\usepackage{rotating}
\restylefloat{figure}
\usepackage{mathrsfs}
\usepackage{fancyhdr}
\usepackage[breaklinks]{hyperref}
\usepackage[numbers,sort&compress]{natbib}
\usepackage{color}
\theoremstyle{plain}
\newtheorem{lem}{Lemma}
\newtheorem{prop}[lem]{Proposition}
\newtheorem{thm}[lem]{Theorem}

\theoremstyle{definition}

\newtheorem{exmp}[lem]{Example}
\newtheorem{rem}[lem]{Remark}
\newcommand{\C}{\mathbb C}
\newcommand{\R}{\mathbb R}
\renewcommand{\S}{\mathbb S}

\newcommand{\Z}{\mathbb Z}
\newcommand{\N}{\mathbb N}
\newcommand{\eps}{\varepsilon}

\newcommand{\dx}{\,\text{\rm d}x}
\renewcommand{\d}{\,\text{\rm d}}
\newcommand{\dw}{\text{\rm d}}
\renewcommand{\phi}{\varphi}
\renewcommand{\theta}{\vartheta}
\newcommand{\norm}[1]{\left|\!\left|#1\right|\!\right|}

\def\lnorm#1{\left|\!\left|{#1}\right|\!\right|}
\newcommand{\ska}[2]{\left\langle #1,#2\right\rangle}
\def\scapro#1#2{\left<#1,#2\right>}
\newcommand{\sigmaess}{\sigma_{\text{\rm ess}}}
\newcommand{\E}{\mathbb E}
\newcommand{\tr}{\text{\rm tr}\;}
\newcommand{\supp}{\text{\rm spt}}
\newcommand{\dist}{\text{\rm dist}\;}
\newcommand{\CalH}{\mathcal{H}}
\newcommand{\CalB}{\mathcal{B}}
\newcommand{\dec}{\text{dec}}
\def\ee{{\text e}}
\def\la{\lambda}
\def\Ran{\text{Ran}}
\def\rom#1{\text{#1}}
\newcommand{\bea}{\begin{eqnarray}}
\newcommand{\eea}{\end{eqnarray}}
\newcommand{\beq}{\begin{equation}}
\newcommand{\eeq}{\end{equation}}
\newcommand{\red}{\color{black}}
\newcommand{\blue}{\color{black}}
\numberwithin{equation}{section}
\numberwithin{lem}{section}
\begin{document}
\title{Bound States for Nano-Tubes with a Dislocation}
\author{Rainer Hempel}
\author{Martin Kohlmann}
\author{Marko Stautz}
\author{J\"urgen Voigt}
\address{Institute for Computational Mathematics, Technische
Universit\"at Braun\-schweig, Pockelsstra{\ss}e 14, 38106
Braunschweig, Germany}
\email{r.hempel@tu-bs.de}
\email{martin.kohlmann@plri.de}
\email{m.stautz@tu-bs.de}
\address{TU Dresden, Fachrichtung Mathematik, Institut f\"ur Analysis, Helmholtzstra{\ss}e 10, 01069 Dresden}
\email{juergen.voigt@tu-dresden.de}
\date{\today}
\begin{abstract} As a model for an interface in solid state physics, we consider two
real-valued potentials $V^{(1)}$ and $V^{(2)}$ on the cylinder or tube $S=\R \times
(\R/\Z)$ where we assume that there exists an interval
$(a_0,b_0)$ which is free of  spectrum of $-\Delta+V^{(k)}$ for $k=1,2$. We are
then interested in the spectrum of $H_t = -\Delta + V_t$, for $t \in \R$, where
$V_t(x,y) = V^{(1)}(x,y)$, for $x > 0$, and $V_t(x,y) =  V^{(2)}(x+t,y)$, for $x < 0$.
 While the essential spectrum of $H_t$ is independent of $t$, we show
 that discrete spectrum, related to the interface at $x = 0$, is created in the interval
$(a_0, b_0)$ at suitable values of the parameter $t$,
provided $-\Delta + V^{(2)}$ has some essential spectrum in $(-\infty, a_0]$.
 We do not require  $V^{(1)}$ or $V^{(2)}$ to be
periodic. We furthermore show that the discrete eigenvalues of $H_t$ are Lipschitz continuous functions of $t$
if the potential $V^{(2)}$ is locally of bounded variation.
\end{abstract}
\maketitle
\section{Introduction}\label{sec_intro}
In the quantum mechanical theory of solids, the spectrum $\sigma(H)$ of self-adjoint Schr\"odinger
operators $H = -\Delta + V$, acting in $\mathsf L_2(\R^3)$, yields basic information on the electronic energy
levels in a crystal. In the most simple cases the potential $V$ is
periodic and then $\sigma(H)$ has \emph{band structure,} i.e., $\sigma(H)$ is the
locally finite union of compact intervals $I_k \subset \R$ with $k \in
\N$. The bands may be separated by (non-empty) open intervals, the
\emph{spectral gaps.} These well-known results are a part of Floquet theory;
cf., e.g., \cite{RS-IV}, \cite{E73}.
%
In reality, however, a solid body occupying all of $\R^3$ in a periodic way is
an idealization and one has to deal with various deviations from
periodicity which may correspond to almost periodic or random potentials.
Notice that there is considerable interest in the properties of ordered, but
non-periodic, solid materials (cf.\ e.g.\ Baake et al.\
\cite{BBM10}, Baake and Grimm \cite{BG10}, Penrose \cite{P79}, Radin \cite{R97}).
In addition, one would like to understand the effect of
\emph{interfaces} occuring in certain types of
alloys where two different structures meet, or of
\emph{surfaces} where a solid only fills a half-space.

In order to facilitate the mathematical analysis of electronic levels
associated with interfaces or surfaces, {\it dislocation potentials}
have been introduced some time ago in the $1$-dimensional case; cf.~the introduction in
\cite{KS12} for the history of this problem.
Here one starts from a periodic, real-valued potential $V = V(x)$ on the real line
and considers the Schr\"odinger operators  $h_t = -\frac{\d^2} {\d x^2} + V_t$, for $t \in \R$,
where $V_t(x) = V(x)$, for $x > 0$, and $V_t(x) = V(x+t)$, for $x < 0$.
Assuming that the spectrum of the periodic operator $h_0$ has a non-trivial gap
$(a_0, b_0)$, located above the infimum of the (essential) spectrum of $h_0$,
one {\blue can show} that $h_t$ has some (discrete)  spectrum in $(a_0,b_0)$ for
suitable $t$.

In the present paper, we will study dislocation problems on an infinite cylinder
$S := \R \times (\R/\Z)$ {\blue without a periodicity assumption.}
 Given two (bounded and measurable) potentials
$V^{(k)} \colon S \to \R$, $k = 1, 2$, the family of {\it dislocation potentials}
is defined by
%
\beq\label{dislocpot}
V_t(x,y)=
\left\{
\begin{array}{ll}
  V^{(1)}(x,y), & x\geq 0, \\
  V^{(2)}(x+t,y), & x<0,
\end{array}
\right.
\eeq
for $(x,y) \in S$ and $t \in \R$.
In the Hilbert space $\mathcal H := \mathsf L_2(S)$, we let
$L$  denote the (unique) self-adjoint extension of $-\Delta$ defined on ${\mathsf C}_{\mathrm c}^\infty(S)$.
For each $t \in \R$ the Schr\"odinger operator $H_t = L + V_t$
describes the energy of an electron on a tube made of the same
or two different materials to the left and to the right of
the interface $\{0\} \times \R/\Z$. We are interested
in the bound states produced by and at this junction where we focus on
energies in a spectral gap of $H_0$. In our main theorem, given below, we will also need
the Dirichlet Laplacian $L_{(0,\infty)}$ of $S^+ := (0,\infty) \times \R/\Z$,
defined as the Friedrichs extension of $-\Delta$ on $\mathsf C^\infty_{\text c}(S^+)$.
\thm\label{ThmI}
Let $V^{(1)}, V^{(2)} \colon S \to \R$ be bounded and measurable, and let $V_t$ be
as in  \eqref{dislocpot}. Suppose $E \in \R$ is such that
\beq\label{cond_0.1}
    E \notin \sigma(L + V^{(k)}),     \qquad k = 1,2,
\eeq
and
\beq\label{cond_0.2}
   \inf \sigmaess(L_{(0,\infty)} + V^{(2)}\! \restriction \! S^+) < E.
\eeq
Then there exists a sequence $(\tau_j)_{j \in \N} \subset [0,\infty)$ of dislocation parameters such that
$E \in \sigma(L + V_{\tau_j})$, and $\tau_j \to \infty$ as $j \to \infty$.
\endthm\rm
\noindent {\it Remarks.}

(a) For the case of periodic potentials $V^{(k)}$ on the real line or on $\R^2$
corresponding results can be found in the paper \cite{HK1}.
The assumptions in Theorem~\ref{ThmI} are purely spectral and do not involve any
further features of the potentials. In this sense, the occurrence of eigenvalues
in gaps for dislocation problems is not an exception, but it is the rule;
 to convey this message is the main objective of the present investigations.

(b) In many applications of Theorem~\ref{ThmI} both $L + V^{(1)}$ and  $L + V^{(2)}$
 have some essential spectrum below the common gap $(a_0,b_0)$.
 In the case of half-space problems, however, only one of the operators
has essential spectrum below the gap; cf.~Example~\ref{half-space-problems}.

(c) We expect the statement of  Theorem~\ref{ThmI} to be true for all $E \in \R$
that satisfy condition \eqref{cond_0.2} and $E \notin \sigmaess(L + V^{(k)})$ for
 $k = 1, 2$.

(d) Our proof of Theorem~\ref{ThmI} is based on an approximation on large sections
$(-n-t,n) \times \R/\Z$ of the tube, much as in \cite{HK1}
where periodic boundary conditions at the ends $-n-t$ and $n$
have been used. Since the potentials $V^{(k)}$ need not be periodic in $x$, there is
no natural boundary condition at the ends, and we simply take Dirichlet boundary
conditions. Of course, this may introduce spurious eigenvalues into the gap and
we adapt a technique of Deift and Hempel \cite{DH86, ADH89} to remove those eigenvalues.
This rather technical construction is at the heart of Section 3.

\vskip1em
 The Laplacian $L$ of Thm.~\ref{ThmI} is unitarily equivalent to the
operator $L_{\text{per}}$, defined as the self-adjoint
realization of $-\Delta$ on the strip $\R \times (0,1)$ with periodic boundary
conditions in $y$. We may extend the potentials $V^{(k)}$, $k=1,2$, and
$V_t$ periodically with respect to the
 $y$-variable to all of $\R^2$, and consider the dislocation problem in
$\R^2$ with the operators $H_t = -\Delta + V_t$. Then Thm.~\ref{ThmI} can
be used to obtain lower bounds for the integrated density of states {\red inside
a gap $(a_0,b_0)$} for certain values of the parameter $t$. A precise description
of the result requires quite a bit of notation and we refer to Section~\ref{sec_exmp}
for details.


We finally address the question of continuity of the (discrete) eigenvalues of
the family of operators $H_t$ as functions of $t$. For periodic potentials in
one dimension continuity is easy as $V_0 - V_t$ tends to zero
{\red in $L_{1, \text{loc,unif}}(S)$, as $t \to 0$; cf.~the appendix in \cite{HK1}}.
Without periodicity,  we now have to face the problem that, no matter how
small $t > 0$ might be, $V_0 - V_t$ need not be small on the global scale.
Here we use a change of variables to the effect that, in the new coordinates,
the potential is altered only in a compact set. This leads to the following
basic
result.
%
\vskip-.5ex
\thm\label{thm_lipschitz}
Let $V^{(1)}, V^{(2)} \in \mathsf L_\infty(S)$ be real-valued, and let
$H_t := L + V_t$ with $V_t$ as in  \eqref{dislocpot}. Then the discrete eigenvalues of $H_t$
depend continuously on $t \in \R$.
If, in addition, the distributional derivative $\partial_1 V^{(2)}$ is a (signed)
Borel measure, 
the discrete eigenvalues of $H_t$ are (locally) Lipschitz continuous
functions of $t \in \R$.
\endthm\rm
%
The second part of the theorem applies in particular if $V^{(2)}$ is of locally bounded
variation{\red ; cf.~\cite{EG92} for details on this class of functions.
Note that Thm.~\ref{thm_lipschitz} also applies to discrete eigenvalues
below the essential spectrum of $H_t$. Corresponding results for the
case of periodic potentials on the real line are given in \cite{HK1}.}

%
%
%
%

\vskip1em


The paper is organized as follows. Section~\ref{sec_prelim} introduces basic notation and describes
the main technical tools used later on. There are several lemmas that will allow us
to control the decoupling via additional Dirichlet boundary conditions. The finer
results depend on exponential decay estimates, which we recall. Another line
of argument deals with coordinate transformations on $S$.

In Section~\ref{sec_main} we construct operators $\tilde H_{n,t}$, acting in
$\mathsf L_2((-n-t,n) \times \R/\Z)$, that serve as an approximation to $H_t$.
We show that there exists a bounded sequence
of parameters $t_n \ge 0$ such that $E$ is an eigenvalue of $\tilde H_{n,t_n}$.
 Taking limits then leads to a $\overline t \ge 0$ such that
$E \in \sigma(H_{\overline t})$ {\red and we may take $\overline t$ as the first
parameter $\tau_1$ in Thm.~\ref{ThmI}.}

Section~\ref{sec_exmp} contains examples of various kinds. The first two examples show
that there exist potentials $V = V(x,y)$ on $S$ that satisfy the assumptions of
Thm.~\ref{ThmI} without requiring  periodicity.
The third example explains in some detail how to use periodic extension in $y$ in order to
go from Thm.~\ref{ThmI} and the tube $S$ to problems in $\R^2$ with potentials
that are periodic in the $y$-variable only.
The fourth example is an application of Thm.~\ref{ThmI} and Thm.~\ref{ThmII} to half-space
problems and the existence of surface states.

In Section~\ref{sec_lipschitz} we analyze continuity properties of the discrete eigenvalues
of $H_t$ as functions of the parameter $t$. We obtain, in particular, Lipschitz continuity
if the distributional derivative $\partial_1 V^{(2)}$ is a measure.

Finally, the appendix gives a detailed account of some basic Hilbert-Schmidt-estimates
on resolvent differences which are then used to control the decoupling by additional Dirichlet
boundary conditions; this material is fairly standard and has been included chiefly
for the convenience of the reader. In addition, we recall a basic method for
obtaining exponential decay estimates for resolvent kernels, {\red and we also give
a proof of Lemma~\ref{lem2.5} which is based on exponential decay.}

We conclude the introduction with some remarks on the literature.
For results concerning the existence of eigenvalues in gaps for dislocation problems in
one dimension, we
 refer to the work of Korotyaev \cite{K00, K05}, Korotyaev and Schmidt \cite{KS12},
Hempel and Kohlmann \cite{HK1}, Dohnal, Plum and Reichel \cite{DPR08}, 
and the references given therein.  While many of the deeper results in 1-d clearly depend on the
 periodicity assumption, the basic question of
\emph{existence} of surface states in a gap can be treated
 in great generality using limit-point/limit-circle theory (cf.~Stautz \cite{St12}).

Several basic questions concerning surface states and the associated integrated
density of states in higher dimensions are discussed in Davies and Simon \cite{DaS78},
Englisch et al.\ \cite{EKSchrS}, and Kostrykin and Schrader \cite{KS01}. However,
there are only few results concerning the {\it existence} of eigenstates in a gap for
higher-dimensional dislocations; cf.\ Hempel and Kohlmann \cite{HK2}.
Dohnal, Plum and Reichel \cite{DPR11} study eigenstates of a non-linear
problem at an interface with energies below the spectrum of the linear operator.
As for the general history of surface states, going back to Rayleigh, Kelvin, and
Landau, we recommend the paper \cite{KhP97} by Khoruzhenko and Pastur.

\section{Preliminaries}\label{sec_prelim}
In this section we introduce notation and collect some basic
results related to the dislocation problem.
\subsection{Notation and Basic Assumptions.}
For basic notation and definitions concerning self-adjoint operators in Hilbert space, we refer
to \cite{K66} and \cite{RS-IV}. The spectral projection associated with a self-adjoint
operator $T$ and an interval $I \subset \R$ is denoted as $\E_I(T)$. If $T$
has purely discrete spectrum in $I$, the number of eigenvalues (counting
multiplicities) in $I$ is given by the trace of $\E_I(T)$, denoted
as $\text{tr }\E_I(T)$.
The Schatten-von Neumann classes will be denoted by $\mathcal B_p$, for $1 \le p < \infty$.

Our basic coordinate space is the tube $S = \R \times (\R/\Z)$ with the usual (flat) product metric,
where $\R/\Z =  \frac1{2\pi} \S^1$. Let us write $\S' := \R/\Z$ for simplicity of notation.
We consider the Sobolev space $\mathsf H^1(S)$ with its canonical norm;
note that $\mathsf C_{\text c}^\infty(S)$ is dense in  $\mathsf H^1(S)$.
Equivalently, we could work with the Sobolev space $\mathsf H^1_{\text{per}}(\R \times (0,1))$
consisting of functions in $\mathsf H^1(\R \times (0,1))$ that are periodic in the $y$-variable.

In the Hilbert space $\mathsf L_2(S)$ we define our basic Laplacian, $L$, to be
the unique (self-adjoint and non-negative) operator associated with the
(closed and non-negative) quadratic form
$$
   \mathsf H^1(S)\ni u \mapsto \int_S|\nabla u|^2\dx\d y,
$$
by the first representation theorem (\cite[Thm.~VI-2.1]{K66}).
As on the real line, the Laplacian $L$ is essentially self-adjoint on
 $\mathsf C_{\text c}^\infty(S)$. $L$ is unitarily equivalent to
the Laplacian $-\Delta$ acting in $\mathsf L_2(\R \times (0,1))$ with
periodic boundary conditions in the $y$-variable.

For $M \subset \R$ open we denote by $L_M$ the Friedrichs extension of $-\Delta$,
defined on $\mathsf C_{\text{c}}^\infty(M \times \S')$, in $\mathsf L_2(M \times \S')$;
in other words, the form domain of $L_M$ is given as the closure of
$\mathsf C_{\text{c}}^\infty(M \times \S')$ in $\mathsf H^1(S)$. Frequently, $M$ will be an open interval
on the real line, or a finite union of such intervals, as in $L_{(\alpha,\beta)}$ for
$-\infty \le \alpha < \beta \le \infty$, or in
$L_{\R \setminus \{\gamma\}} = L_{(-\infty,\gamma)} \oplus L_{(\gamma,\infty)}$ for $\gamma \in \R$.
If $M = (\alpha,\beta)$ for some $-\infty < \alpha < \beta < \infty$, we say that
$L_{(\alpha,\beta)}$ satisfies Dirichlet boundary conditions on the lines
$\{\alpha\} \times \S'$ and $\{\beta\} \times \S'$.

%
Given two bounded, measurable functions $V^{(1)}$, $V^{(2)}\colon S\to\R$ we
define the Schr\"odinger operators $H^{(k)}=L+V^{(k)}$, for $k=1,2$.
Throughout Sections~\ref{sec_prelim} and \ref{sec_main},
we assume $V^{(k)} \ge 0$ for simplicity (and without loss of
generality). For $t\in\R$ the dislocation potentials
$V_t$ are defined as in \eqref{dislocpot}, and we let $H_t=L+V_t$, $t \in \R$,
denote the family of dislocation operators.

From a technical point of view, the following three tools are fundamental for our approach:

\begin{itemize}
\item decoupling by Dirichlet boundary conditions on circles $\{c\} \times \S'$,
\item exponential decay of eigenfunctions,
\item a coordinate transformation with respect to the $x$-variable.
\end{itemize}
We provide some preliminary facts concerning these tools here.
We begin with Dirichlet decoupling.

\subsection{Dirichlet Decoupling.}
In this subsection we show how to control the effect of an additional Dirichlet
boundary condition on {\red the line $\{0\} \times \S' \subset S$; topologically,
 $\{0\} \times \S' \subset S$ is a circle.
On the strip $\R \times (0,1)$ with periodic boundary conditions the additional
Dirichlet boundary condition would be placed on the straight line segment
$\{0\} \times (0,1)$. Note that it is essential for our applications later on to have}
estimates with constants that are uniform for certain classes of potentials.
\lem\label{lem2.1} Let $0 \le W \in \mathsf{L}_\infty(S)$, let
$H = L + W$ in the Hilbert space $\CalH = \mathsf L_2(S)$,
 and let $H_D := L_{\R \setminus \{0\}} + W$.

Then $(H + r)^{-1} - (H_D + r)^{-1}$ is Hilbert-Schmidt for all $r \ge 1$
and there is a constant $C \ge 0$, which is independent of $W$ and $r$, such that
\beq
   \lnorm{(H + r)^{-1} - (H_D + r)^{-1}}_{\CalB_2(\CalH)} \le C,
   \qquad r \ge 1.
\label{eqlem2.1}\eeq
\endlem\rm
Estimates of type \eqref{eqlem2.1} are well-known and have been of great use in spectral and in
scattering theory. For the convenience of the reader, we give a sketch of the
proof in the appendix. Similar estimates hold for finite tubes $(-n,n) \times \S'$ where we compare
$L_{(-n,n)}$ and $L_{(-n,n) \setminus \{0\}} = L_{(-n,0)} \oplus L_{(0,n)}$.
\lem\label{lem2.2} Let $0 \le W \in \mathsf{L}_\infty(S)$ and let $L_{(-n,n)}$ and $ L_{(-n,n) \setminus \{0\}}$
be as above. Then  $(L_{(-n,n)} + W + r)^{-1} - ( L_{(-n,n) \setminus \{0\}} + W + r)^{-1}$ is
Hilbert-Schmidt for $r \ge 1$ and we have an estimate
$$
   \lnorm{(L_{(-n,n)} + W + r)^{-1} - ( L_{(-n,n) \setminus \{0\}} + W + r)^{-1}}_{\CalB_2(\CalH)} \le C,
$$
with a constant $C$ independent of $r$ and $W$.
\endlem\rm
Indications on the proof are given in the appendix.

It is easy to generalize the above results to situations where
we add in Dirichlet boundary conditions on several lines of the type
$\{x_0 \} \times \S'$. This immediately gives a simple proof for the invariance
of the essential spectrum. 
\prop\label{prop2.3}  For $k = 1,2$, let $V^{(k)} \in  \mathsf{L}_\infty(S)$
and define $H^{(k)}$ and $H_t$ as above. In addition, let $H^{(1)}_+ :=
L_{(0,\infty)} + V^{(1)}$ and  $H^{(2)}_- := L_{(-\infty,0)} + V^{(2)}$.
We then have
\beq
   \sigmaess(H_t) = \sigmaess(H^{(1)}_+) \cup  \sigmaess(H^{(2)}_-)
   \subset  \sigmaess(H^{(1)}) \cup  \sigmaess(H^{(2)}), \qquad
   t \in \R.
\label{eqprop2.3}
\eeq
\endlem\rm
\proof {\red (cf.\ also \cite{HK1})} 
For $t \ge 0$, let $H_{t,\dec}$ denote the operator obtained from $H_t$ by
the insertion of Dirichlet boundary conditions on the lines $\{0\}\times \S'$ and
 $\{-t\}\times \S'$. By Lemma~\ref{lem2.1}, $(H_t + 1)^{-1} - (H_{t,\dec} + 1)^{-1}$
is compact and so $H_t$ and $H_{t,\dec}$ have the
same essential spectrum. The part of $H_{t,\dec}$ to the left of $-t$ is unitarily equivalent
to $H^{(2)}_-$, and the part of $H_{t,\dec}$ associated
with the interval $(-t,0)$ has compact resolvent. Thus $\sigmaess(H_{t,\dec})
= \sigmaess(H_{0,\dec}) =  \sigmaess(H^{(1)}_+) \cup  \sigmaess(H^{(2)}_-)$.
This proves the equality in  \eqref{eqprop2.3}. The inclusion stated in \eqref{eqprop2.3} is
immediate from Lemma~\ref{lem2.1}.
\endproof
The result of Prop.~\ref{prop2.3} can also be obtained by considering singular
sequences (so-called {\it Weyl-sequences}).
Yet another method of proof, based on a transformation of coordinates,
is given in Section~\ref{subsec_trafo} below.
\subsection{Exponential Decay of Eigenfunctions.}
The following contains our basic exponential decay estimate.
The proof uses classic arguments as discussed in
\cite{S82} or \cite{HS96}; cf.~also
\cite{vdBHV} for a recent version. It is of importance for the applications
we are having in mind that the bound of Lemma~\ref{lem2.4} below is independent of $W$
within the class of bounded, non-negative potentials with a given
spectral gap $(a_0, b_0)$.
We let $\chi_k$ denote the characteristic function of the
set $[-k,k]\times \S' \subset S$, for brevity. Again, the proofs of the statements
in this subsection are deferred to the appendix.
\lem\label{lem2.4} For $0 \le a_0 < a < b < b_0$ given there exist constants
$C \ge 0$ and $\gamma > 0$ such that for all $0 \le W \in\mathsf{L}_\infty(\R)$
with $\sigma(L + W) \cap (a_0,b_0) = \emptyset$ we have
$$
     \lnorm{(1 - \chi_k) u} \le C \ee^{-\gamma k} \lnorm{u}, \qquad k \in \N,
$$
for all eigenfunctions $u$ of $L_{\R \setminus \{0\}} + W$ that are associated
with an eigenvalue $\lambda \in [a,b]$.
\endlem\rm
The following lemma gives an upper bound for the number of eigenvalues that are
moved into a compact subset $[a,b]$ of a spectral gap $(a_0,b_0)$ upon
enforcing a Dirichlet boundary condition on the
line $\{0\}\times \S'$. Again, it is important
that the bound is {\it independent} of the potential $W$, provided $W \ge 0$.
A proof is given in the Appendix.
\lem\label{lem2.5} For numbers $a_0 < a < b < b_0 \in \R$ given there exists
a constant $c \ge 0$ with the following property: If $0 \le W \in\mathsf{L}_\infty(S ; \R)$ satisfies
$\sigma(L+W) \cap (a_0,b_0) = \emptyset$, then
$$
    \tr \E_{[a,b]}(L_{\R \setminus \{0\}} + W) \le c.
$$
\endlem\rm
\subsection{Transformation of Coordinates.} \label{subsec_trafo}

Some additional insight can be gained by using a transformation of coordinates which,
in a sense, ``undoes'' the effect of the dislocation outside a finite
section of the tube $S$. In this way, the dislocation problem can be viewed as a
perturbation which acts in a compact subset of $S$ only.
To this end, we provide (smooth) diffeomorphisms $\phi_t \colon \R \to \R$
of class $\mathsf C^\infty$ with the additional properties that
$$
    \phi_t(x) = x, \quad x \ge 0, \qquad \phi_t(x) = x - t, \quad x \le -2;
$$
we also require that there is a constant $C \ge 0$ s.th.\
$$
   \max_{x\in\R} |\varphi_t(x) - x|, \quad
  \max_{x\in\R} |\varphi'_t(x) - 1|, \quad
   \max_{x\in\R} |\varphi_t''(x)| \le C t, \qquad t \in [0,2].
$$
In Section~\ref{sec_lipschitz} it will be shown that, for $0 \le t \le 1$, the dislocation
operators $H_t$ are unitarily equivalent to (s.a.)~operators ${\hat H}_t$
acting in $\mathsf L_2(S)$ with domain $D({\hat H}_t) = D(L)$ where the
quadratic form of  ${\hat H}_t$ is given by
\begin{align}
  \hat H_t[u,u] & :=
 \int_{S} \left(
                     \frac{1}{(\varphi_t')^2} |\partial_1 u |^2
                + |\partial_2 u|^2
     -  \frac{\varphi_t''}{(\varphi_t')^3}\, \hbox{\rm Re}\, ({\bar u}  \partial_1 u )
     +    \frac{(\varphi_t'')^2 }{ 4  (\varphi_t')^4} |u|^2
             \right)\dw x\d y \cr
     & \qquad \qquad \qquad +  \int_{S} V_t(\phi_t(x),y) |u|^2\dx\d y;
\nonumber
\end{align}
here ${\hat H}_0 = H_0 = H$.
Note that ${\hat H}_t - H_0$ has support in the compact set $\{-2 \le x
\le 0 \}$. In other words, the family $({\hat H}_t)_{0 \le t \le 1}$ gives an
equivalent description of the dislocation problem where the perturbation
is now restricted to the bounded set $\{ (x,y) \in S \mid -2 \le x  \le 0 \}$.

The family $({\hat H}_t)_{0 \le t \le 1}$ enjoys the following properties:

\noindent (1) The mapping $[0,1] \ni t \mapsto ({\hat H}_t + 1)^{-1}$ is norm-continuous.

\noindent (2) For $t, t' \in [0,1]$, the resolvent difference $({\hat H}_t + 1)^{-1}
  - ({\hat H}_{t'} + 1)^{-1}$ is compact.

Both properties will be proved in Section~\ref{sec_lipschitz}. It follows from (2) that the
essential spectrum of ${\hat H}_t$ is stable, and then the same property holds
for the family $(H_t)_{0 \le t \le 1}$; cf.~Prop.~\ref{prop2.3}.
Property (1) implies that the spectrum of ${\hat H}_t$ depends
continuously on $t$ in the usual Hausdorff-metric on the
real line, and then the same holds for the family $(H_t)_{0 \le t \le 1}$.
\section{The main result}\label{sec_main}
In this section we give a proof of Theorem~\ref{ThmI}.
We consider some $E \in \R$ satisfying the
assumptions \eqref{cond_0.1} and \eqref{cond_0.2} of Thm.~\ref{ThmI}.
It follows from condition \eqref{cond_0.1} that there is an $\alpha > 0$ such that
%
$$     \dist(E, \sigma(L + V^{(k)})) \ge 2\alpha, \qquad k = 1,2;$$
%
$E$ and $\alpha$ will be kept fixed throughout this section. If it happens that
$E$ is an eigenvalue of $H_0 = L + V_0$ we set $\tau_1 := 0$ and consider $H_1$ instead of
$H_0$. We may therefore assume in the sequel that $E
\notin\sigma(H_0)$. We now fix some $0 < \beta \le 2\alpha/3$ such that
\beq\label{distance_E_sp(H0)}
      \dist(E, \sigma(H_0)) \ge 3\beta.
\eeq
We find solutions of suitable approximating problems, and then pass
to the limit. The basic idea is to restrict the problem to finite sections of
the tube $S$ of the form $(-n-t, n) \times \S'$, as in \cite{HK1, HK2} where $S$ is a
strip and the potential $V$ is periodic. {\red In \cite{HK1, HK2} periodic boundary conditions
at the ends of the finite strip work nicely, but for non-periodic potentials
there is no natural choice of boundary conditions
on the lines $\{\pm n\} \times \S'$ that would keep the interval
$(E - \beta, E + \beta)$  free of spectrum of the operators
$H_{n,0} = L_{(-n,n)} + V_0$ and we have to resort to a more complicated construction,
 inspired by some work of Deift and Hempel \cite{DH86} (cf.~also \cite{ADH89}}).

\subsection{The Approximating Problems.}

We first introduce  ``correction terms'' in the form of projections, sandwiched between
suitable cut-offs. While we have two interacting Dirichlet boundaries, we prefer a construction
where the correction term at the left end does not depend on the correction term
at the right end. Let
\begin{align}
\begin{split}
  H_n^+=L_{(-\infty,n)}+V^{(1)} & \text{ in }\mathsf L_2((-\infty,n)\times \S'),
   \\
  H_n^-=L_{(-n,\infty)}+V^{(2)} & \text{ in }\mathsf
  L_2((-n,\infty)\times \S'),
\nonumber
\end{split}
\end{align}
{\red for $n \in \N$, where we have chosen the upper indices $\pm$ of $H_n^\pm$ in reference to the Dirichlet
boundary condition on the lines $\{\pm n\} \times \S'$. As in Prop.~\ref{prop2.3}, we have
$\sigmaess(H_n^+) \subset \sigmaess(H^{(1)})$ and $\sigmaess(H_n^-) \subset \sigmaess(H^{(2)})$ so
that $(E - 3\beta, E + 3\beta)$ is a gap in the essential spectrum of
$H_n^\pm$.}
%
{\red We are now going to construct a family of operators ${\tilde H}_{n,t}$ on $(-n-t,n) \times \S'$
that will serve as approximations to $H_t$ and which enjoy the property that the interval
$(E-\beta,E+\beta)$ is free of spectrum of  ${\tilde H}_{n,0}$. }

Let $\Phi_{n,k}^\pm$, $k = 1, \ldots, J_n^\pm$, denote a (maximal) orthonormal
set of eigenfunctions of $H_n^\pm$ corresponding to its eigenvalues in $[E-2\beta, E+2\beta]$.
By Lemma~\ref{lem2.5}, there is a constant $c_0$
 such that $J_n^\pm\leq c_0$ for all $n$; {\red here we apply Lemma 2.5 twice, with the choice
$a_0 := E - 3\beta$, $a := E - 2\beta$, $b := E + 2\beta$, $b_0 := E +
3\beta$, and $W := V^{(1)}(. + n)$ or  $W := V^{(2)}(. - n)$, respectively. }

Next, we introduce the projections $P^\pm_n$ onto the span of the $\Phi_{n,k}^\pm$, given by
%
$$P_n^\pm = \E_{[E-2\beta, E+2\beta]}(H_n^\pm).$$
As a consequence,
\beq\label{specHnpm}\sigma(H_n^\pm+4\beta P_n^\pm)\cap[E-2\beta,E+2\beta]=\emptyset.\eeq
 Here the eigenfunctions $\Phi_{n,k}^\pm$ are localized near $\{\pm n\} \times
 \S'$ and decay (exponentially) as
$x$ increases or decreases from $\pm n$, cf.~Lemma~\ref{lem2.4}. We are now
going to make this more precise.

{\red Let us first introduce some cut-off functions.} Let $\chi_1^+\in \mathsf{C}^\infty(-\infty,1)$
 with $0\leq\chi_1^+\leq 1$, $\chi_1^+(x)=1$ for $x>3/4$, and $\chi_1^+(x)=0$
 for $x<1/2$, be given.
Now set $\chi_n^+(x)=\chi_1(x/n)$, so that $\chi_n^+\in \mathsf{C}^\infty(-\infty,n)$,
$\chi_n^+(x)=1$ for
 $x>3n/4$, and $\chi_n^+(x)=0$ for $x<n/2$. We define $\chi_n^-\in \mathsf{C}^\infty(-n,\infty)$ analogously
by setting $\chi_n^-(x)=\chi_n^+(-x)$.
Furthermore, choose $\phi_1\in \mathsf{C}_{\text{c}}^\infty(-1/2,1/2)$ with $0\leq\phi_1\leq 1$ and $\phi_1(x)=1$ for $|x|<1/4$,
and set $\phi_n(x)=\phi_1(x/n)$ and $\psi_n=1-\phi_n$.
Finally, we decompose $\psi_n=\psi^-_n+\psi^+_n$ and note that $\psi_n^\pm\chi_n^\pm=\chi_n^\pm$.
{\red By Lemma~\ref{lem2.4}} there are constants $c \ge 0$ and $n_0 \in \N$ such that
%
$$\norm{ ( 1 - \chi_n^\pm) \Phi_{n,k}^\pm} \le c/n,
   \qquad   n \ge n_0,$$
and we infer that {\red there is a constant $C \ge 0$ such that }
\beq\label{limproj}\norm{\chi_n^\pm P_n^\pm\chi_n^\pm-P_n^\pm}\leq\frac{C}{n};\eeq
%
in fact, a stronger
estimate of the form  $\norm{\chi_n^\pm P_n^\pm\chi_n^\pm-P_n^\pm}\leq
C' e^{-\gamma n}$, for some $\gamma > 0$, holds true. We now define
%
$$\tilde H_n^\pm=H_n^\pm+4\beta\chi_n^\pm P_n^\pm\chi_n^\pm$$
and observe that, by (\ref{specHnpm}) and (\ref{limproj}),
%
$$\sigma(\tilde H_n^\pm)\cap[E-\beta,E+\beta]=\emptyset, $$
for $n$ large.
In particular, for any $u\in D(\tilde H_n^\pm)=D(H_n^\pm)$ we have
\beq\label{estimate_alpha}\norm{u}\leq\frac{1}{\beta}\norm{(\tilde H^\pm_n-E)u}.\eeq
Now the dislocation enters the game: let $T_t(x,y)=(x+t,y)$, for $(x,y) \in
S$,  and define
%
$$P_{n,t}^-=\sum_{k\in J_n^-}\ska{\, . \, }{\Phi_{n,k}^-\circ T_t}\Phi_{n,k}^-\circ T_t,$$
as well as $\chi^-_{n,t} := \chi^-_n \circ T_t$.
Finally, let
%
$$
 \mathcal P_{n,t}
     =4\beta \left( \chi_n^+P_n^+\chi_n^+ + \chi_{n,t}^- P_{n,t}^-\chi_{n,t}^- \right)
$$
and
%
$$\tilde H_{n,t}=L_{(-n-t,n)}+V_t+\mathcal P_{n,t}
$$
in $\mathsf L_2((-n-t,n)\times \S')$.
The operators ${\tilde H}_{n,t}$ are the principal players in our
approximating problems. We first establish that the operators ${\tilde H}_{n,0}$
have no spectrum in the interval $[E - \beta, E + \beta]$, for $n$ large.
%
\vskip1em

\lem\label{lem3.1} Let $E \in \R \setminus \sigma(H_0)$ satisfy condition
\eqref{cond_0.1} of Thm.~\ref{ThmI}
 and let $\beta$ be as in  \eqref{distance_E_sp(H0)}.
Then there is an $n_0 \in \N$ such that
$$
     \sigma(\tilde H_{n,0}) \cap (E-\beta,E+\beta) = \emptyset,
     \qquad n \ge n_0.
$$
\endlem\rm
\proof Else there exists a sequence $n_j \to \infty$ and there
exist $E_j \in [E-\beta,E+\beta]$ such that $E_j$ is an eigenvalue of
$\tilde H_{n_j,0}$, for $j\in\N$. Let $u_{n_j}$ denote an associated
normalized eigenfunction. With the cut-off functions $\phi_k$
and $\psi_k^\pm$ defined above, we see that
 $\phi_{n_j/4} u_{n_j}  \in D(H_0)$ and $\psi_{n_j/4}^\pm u_{n_j} \in D(\tilde H_{n_j}^\pm)$
with estimates
\beq\label{estimate_psi_pm}
  \norm{(H_0 - E_j)(\phi_{n_j/4} u_{n_j}) } \le c/n_j,
  \qquad
   \norm{(\tilde H_{n_j}^\pm - E_j)(\psi^\pm_{n_j/4} u_{n_j}) } \le c/n_j,
\eeq
for $j$ large; here $c \ge 0$ is a suitable constant.
 Since $\sigma(H_0) \cap (E - 2\beta, E + 2\beta) = \emptyset$ and
$E_j \in [E-\beta, E+\beta]$, we have
$\norm{(H_0 - E_j)u} \ge \beta \lnorm{u}$ for all $u \in D(H_0)$,
 so that $\phi_{n_j/4} u_{n_j} \to 0$ as $j \to \infty$.
Similarly, the second estimate in \eqref{estimate_psi_pm} and  \eqref{estimate_alpha} imply
 that $\psi_{n_j/4}^\pm u_{n_j} \to 0$, as $j \to \infty$.
Therefore $u_{n_j} \to 0$ as $j\to \infty$, in contradiction to $\norm{u_{n_j}} = 1$.
\endproof

\subsection{Solution of the Approximating Problems.}

We are now going to show that, for large $n \in \N$,
there exist parameters $t_n \ge 0$ such that $E$ is an eigenvalue of
${\tilde H}_{n,t_n}$. Since all the operators involved have
purely discrete spectrum we can use a simple eigenvalue counting argument.
\prop\label{solution_of_approximating_problems} Let $E \in \R \setminus \sigma(H_0)$
satisfy conditions \eqref{cond_0.1} and \eqref{cond_0.2} of Theorem \ref{ThmI}.
  Then  there are $n_0 \in \N$ and $\gamma_0 > 0$
  such that for any $n \in \N$ with $n \ge n_0$ there exists $ 0 < t_n \le \gamma_0$ such
  that  $E$ is an eigenvalue of ${\tilde H}_{n,t_n}$.
\endlem\rm

In preparation for the proof, we introduce variants of our
operators with  Dirichlet boundary conditions on suitable
lines.  Let ${\tilde H}_{n,t;\text{dec}}$ denote the operator ${\tilde H}_{n,t}$
with additional DBCs on the lines $\{0\} \times \S'$ and $\{-t\}
\times \S'$; note
that---by virtue of the cut-offs $\chi^+_n$ and $\chi_{n,t}^-$---the non-local operators
$\mathcal P_{n,t}$ are not affected by these boundary conditions.
For $n \ge n_0(t)$ the operators ${\tilde H}_{n,t;\text{dec}}$ can be written as direct sums
$$
  {\tilde H}_{n,t;\text{dec}} = {\tilde h}_{n,t;1} \oplus {h}_{t;2} \oplus {\tilde h}_{n;3},
$$
with
$$
 {\tilde h}_{n,t;1}
   : = L_{(-n-t, -t)} + V_t + 4 \beta \chi_{n,t}^- P_{n,t}^- \chi_{n,t}^-,
 $$
acting in $\mathsf L_2((-n-t, -t) \times \S')$ with DBCs on $\{-n-t\} \times \S'$ and on
$\{-t\} \times \S'$,
$$
   h_{t;2} : = L_{(-t,0)} + V_t 
$$
acting in $\mathsf L_2((-t,0) \times \S')$ with DBCs  on $\{-t\}  \times \S'$ and on
$\{0\}  \times \S'$, and, finally,
$$
  {\tilde h}_{n;3} : = L_{(0,n)} + V^{(1)} + 4 \beta  \chi_n^+ P_n^+ \chi_n^+,
$$
acting in $\mathsf L_2((0,n) \times \S')$ with DBCs on $\{0\}  \times \S'$ and on
$\{n\}  \times \S'$.

We now collect some properties of the operators $ {\tilde H}_{n,t; \text{dec}}$
that we need in the proof of Proposition \ref{solution_of_approximating_problems}.

(1) For $t=0$ we have
  \begin{align}
  {\tilde H}_{n,0; \text{dec}} &= {\tilde h}_{n,0;1} \oplus {\tilde h}_{n;3} \nonumber\\
   &  = (L_{(-n,0)} + V^{(2)} +  4 \beta  \chi_n^- P_n^- \chi_n^-)
    \oplus   (L_{(0,n)} + V^{(1)} + 4 \beta  \chi_n^+ P_n^+ \chi_n^+). \nonumber
  \end{align}
{\red The following lemma compares} the number of eigenvalues below $E$ for the
operators ${\tilde H}_{n,0}$ and ${\tilde H}_{n,0; \text{dec}}$.
\lem\label{lem3.3} Let $E$ and $\beta$ satisfy \eqref{distance_E_sp(H0)},
let $\tilde H_{n,0}$ and ${\tilde H}_{n,t;\text{\rm dec}}$ be as above,
and let $n_0$ as in Lemma \ref{lem3.1}.
Then there is a constant $c_0 \ge 0$ such that
$$
    \tr \E_{(-\infty,E]} ({\tilde H}_{n,0}) \ge
   \tr \E_{(-\infty,E]} ({\tilde H}_{n,0; \text{\rm dec}}) \ge
   \tr \E_{(-\infty,E]} ({\tilde H}_{n,0}) - c_0, \qquad  n \ge n_0.
$$
\endlem\rm

In the proof of Lemma \ref{lem3.3} we use a proposition,
based on the Birman-Schwinger principle {\red (cf.~\cite{RS-IV}, \cite{S05})}
 to control the spectral shift across $E$, produced
by the Dirichlet boundary condition on $\{0\} \times \S'$.
 Recall that $\CalB_p$ denotes the $p$-th Schatten-von Neumann class, for $1 \le p < \infty$.
\prop\label{spectral_shift}
Let $1 \le T \le S$ be self-adjoint operators with compact resolvent
in the Hilbert-space $\CalH$, and suppose that $T^{-1} - S^{-1} \in \CalB_p(\CalH)$
for some $p \in [1,\infty)$. Then for any {\red $E \in \R \setminus \sigma(T)$} we have
$$
    \tr \E_{(-\infty,E)}(S) \ge \tr  \E_{(-\infty,E)}(T) - \text{\rm dist}(E,\sigma(T))^{-p}
      \lnorm{T^{-1} - S^{-1}}_{\CalB_p}^p.
$$
\endprop\rm
\proof The proof is immediate from Proposition 1.1 in \cite{H92} with
$A := (T+1)^{-1}$, $B := (S+1)^{-1}$, and $\eta := (E + 1)^{-1}$.
\endproof

\proof[Proof (of Lemma \ref{lem3.3})] The first inequality follows immediately
from ${\tilde H}_{n,0; \text{\rm dec}} \ge {\tilde H}_{n,0}$. To prove the
second inequality, we apply Prop.~\ref{spectral_shift} with
$T := {\tilde H}_{n,0} + 1$, $S := {\tilde H}_{n,0,\dec} + 1$, and $p=2$.
Here $(H_{n,0} + 1)^{-1} - (H_{n,0,\dec} + 1)^{-1}$ is Hilbert-Schmidt
by Lemma~\ref{lem2.2} with a bound $c_1$ on the HS-norm which is independent
of $n$.
Simple perturbational arguments (\cite[Lemma 1.4]{H92}) yield that there exists
a constant $c_2 \ge 0$ such that
$$
    \lnorm{(\tilde{H}_{n,0} + 1)^{-1} - (\tilde{H}_{n,0,\dec} + 1)^{-1}}_{\CalB_2} \le c_2, \qquad n \ge n_0.
$$
Now Prop.~\ref{spectral_shift} implies
$$
 \tr \E_{(-\infty,E)}(\tilde{H}_{n,0,\dec})
    \ge \tr \E_{(-\infty,E)}(\tilde{H}_{n,0}) - \beta^{-2} c_2^2;
$$
here the left hand side is enlarged if we replace $\E_{(-\infty,E)}$ with  $\E_{(-\infty,E]}$ while the
right hand side remains unchanged under this replacement since $E \notin \sigma(\tilde{H}_{n,0})$ .
 \endproof
%

(2) The operator  $\tilde h_{n,t;1}$ is unitarily equivalent to $\tilde h_{n,0;1}$ via a right-translation
 through $t$ so that
\beq\label{nt1n0dec}
 \tr \E_{(-\infty,E]} ({\tilde h}_{n,t;1}) + \tr \E_{(-\infty,E]} ({\tilde
   h}_{n ;3})
  = \tr \E_{(-\infty,E]} ({\tilde H}_{n,0; \text{\rm dec}}).
 \eeq

 (3) The operators $h_{t;2}$ are unitarily equivalent to
$L_{(0,t)} + V^{(2)} \restriction (0,t)$ for all $t > 0$ by a right translation and
we have the following lemma.
%
%
%
\lem\label{lem_3.5} Let  $h_{t;2}$ as above and let $E$ and $V^{(2)}$ satisfy
condition \eqref{cond_0.2}.
Then
%
$$
       \tr \E_{(-\infty, E)} (h_{t;2})
    \to \infty, \qquad t \to \infty.
$$
\endlem\rm

For the proof we prepare a lemma.
\lem\label{lem_3.6}
Let $A$ and $A_n$, for $n \in \N$,  be bounded, symmetric operators in some Hilbert
space and suppose that $A_n \to A$ strongly. 
Then, for any $\la_0 \in \sigmaess(A)$ and any $\eps > 0$
we have $\tr \E_{(\la_0 - \eps, \la_0 + \eps)}(A_n) \to \infty$ as $n \to \infty$.
\endlem\rm

In Lemma~\ref{lem_3.6} we allow for $\tr \E_{(\la_0 - \eps,\la_0 + \eps)}(A_n) = \infty$; a
precise statement would read as follows: For any $k \in \N$ there exists $n_0 \in \N$
such that $\tr\E_{(\la_0 - \eps,\la_0 + \eps)}(A_n) \in [k, \infty]$ for all $n \ge n_0$.

\proof Assume for a contradiction that there exist
{\red $\la_0 \in \sigmaess(A)$, $k_0 \in \N$,} and a
sequence $(n_j) \subset \N$ with $n_j \to \infty$, as $j \to \infty$,
such that $\tr \E_{(\la_0 - \eps,\la_0 + \eps)}(A_{n_j}) \le k_0$  for all $j \in \N$.

Let $0 < \eps' < \eps$ and choose a continuous function
$f \colon \R \to [0,1]$ such that $f(x) = 1$ for $|x - \la_0| \le \eps'$
and $f(x) = 0$ for $|x - \la_0| \ge \eps$.
By routine arguments, it follows from the assumptions that
$p(A_n) \to p(A)$ strongly for all real-valued polynomials and then
that $f(A_n) \to f(A)$ strongly; here we also use that
the norms $\norm{A_n}$ form a bounded sequence.

There exists an ONS $\{u_1, \ldots, u_{k_0 + 1} \} \subset \Ran\,\E_{(\la_0 - \eps',\la_0 + \eps')}(A)$.
As $\chi_{(\la_0 - \eps, \la_0 + \eps)} \ge f \ge 0$, monotonicity of the trace yields
$$
  \tr\E_{(\la_0-\eps,\la_0 + \eps)}(A_{n_j}) \ge
  \tr f(A_{n_j}) \ge \sum_{m=1}^{k_0 + 1} \scapro{f(A_{n_j})u_m}{u_m}
$$
with $\sum_{m=1}^{k_0 + 1} \scapro{f(A_{n_j})u_m}{u_m} \to
   \sum_{m=1}^{k_0 + 1} \scapro{f(A)u_m}{u_m} = k_0 + 1$, as $j \to \infty$.
\endproof
\proof[Proof (of Lemma~\ref{lem_3.5}).] Since $h_{t;2}$ and $L_{(0,t)} + V^{(2)}$ are
unitarily equivalent, we only have to show
that $ \tr \E_{(-\infty,E)}(L_{(0,t)} +  V^{(2)}) \to \infty$ as $t \to \infty$.
Here we may use Lemma~\ref{lem_3.6}, applied to the operators
$$
   A := (L_{(0,\infty)} + V^{(2)} + 1)^{-1},
 \qquad A_t :=  (L_{(0,t)} + V^{(2)} + 1)^{-1} \oplus 0,
$$
with the operator $0$ acting in $\mathsf L_2((t,\infty) \times \S')$.
Indeed, it follows from a result in \cite{S78}
 that $A_t \to A$ strongly, as $t \to \infty$.
\endproof

Let us note as an aside that there is a kind of converse to the
statement of Lemma~\ref{lem_3.5}:
If $\eta <  \inf \sigmaess(L_{(0,\infty)} + V^{(2)})$,
then min-max and $L_{(0,t)} + V^{(2)} \ge L_{(0,\infty)} + V^{(2)}$ imply that
$$
  \tr\E_{(-\infty,\eta]}(L_{(0,t)} + V^{(2)}) \le \tr\E_{(-\infty,\eta]}(L_{(0,\infty)} + V^{(2)}) < \infty,
  \qquad t > 0,
$$
and thus $\tr\E_{(-\infty,\eta]}(L_{(0,t)} + V^{(2)})$ is a bounded function of $t > 0$.
%

We are now ready for the proof of Proposition~\ref{solution_of_approximating_problems}.
\proof[Proof of Prop.~\ref{solution_of_approximating_problems}.]
Let $E \in (a,b) \setminus \sigma(H_0)$. By Lemma \ref{lem3.1} there exist $\beta > 0$ and
$n_0 \in \N$ such that
$$
   (E - \beta, E + \beta) \cap \sigma(\tilde H_{n,0}) = \emptyset, \qquad n \ge n_0.
$$
Adding in Dirichlet boundary conditions raises eigenvalues and we thus have
\begin{align}
  \tr \E_{(-\infty,E]}({\tilde H}_{n,t})
  & \ge  \tr \E_{(-\infty,E]}({\tilde H}_{n,t;\text{dec}}) \nonumber \\
  & = \tr \E_{(-\infty,E]}({\tilde h}_{n,t;1}) +  \tr
  \E_{(-\infty,E]}(h_{t;2})
              +  \tr \E_{(-\infty,E]}(\tilde h_{n;3}) \nonumber\\
  & =  \tr \E_{(-\infty,E]}(\tilde H_{n,0;\text{dec}}) +  \tr
  \E_{(-\infty,E]}(h_{t;2}),
 \nonumber
\end{align}
where we have used \eqref{nt1n0dec} in the last step. It now follows from Lemma \ref{lem3.3} that
$$
    \tr \E_{(-\infty,E]}({\tilde H}_{n,t}) \ge  \tr \E_{(-\infty,E]}({\tilde H}_{n,0})
      - c_0 +  \tr \E_{(-\infty,E]}(h_{t;2}),
$$
with the constant $c_0$ from Lemma \ref{lem3.3}. Since $V^{(2)}$ satisfies
condition \eqref{cond_0.2},  Lemma \ref{lem_3.5} implies that there exists
$\gamma_0 > 0$  such that $ \tr \E_{(-\infty,E]} (h_{\gamma_0;2}) > c_0$ and
we conclude that
\beq\label{n_t_0_n_0}
    \tr \E_{(-\infty,E]}({\tilde H}_{n,\gamma_0}) >  \tr \E_{(-\infty,E]}({\tilde H}_{n,0}), \qquad
    n \ge n_0.
\eeq
The operators $\tilde H_{n,t}$ have purely discrete spectrum and their eigenvalues depend
continuously on $t \ge 0$, as can be easily seen by arguments similar to the ones
used at the end of Section 2. Therefore,  \eqref{n_t_0_n_0} implies that at
least one eigenvalue of  $\tilde H_{n,t}$ has crossed $E$ at some $0 < t_n \le \gamma_0$,
and we are done.
\endproof
\subsection{Convergence of the Approximative Solutions.}

The above Prop.~\ref{solution_of_approximating_problems} shows that
there exists a bounded sequence of parameters $t_n$ such that
$E$ is an eigenvalue of $\tilde H_{n,t_n}$. Then there is a convergent
subsequence $t_{n_j} \to \bar t$, as $j \to \infty$, and we
expect that $E$ is an eigenvalue of $H_{\bar t}$.
\lem\label{lem_3.7} Suppose we are given sequences $(t_n) \subset [0, \infty)$
and $(E_n) \subset [E- \beta, E + \beta]$ with $t_n \to \overline t$ and $E_n \to E$,
as $n \to \infty$,  with the property that $E_n$ is an eigenvalue of ${\tilde H}_{n,t_n}$ for $n \ge n_0$.
Then $E$ is an eigenvalue of $H_{\overline t}$.
\endlem\rm\vspace{-.5cm}
{\red
\proof  Let $f_n \in  D(H_{n,t_n})$ be normalized eigenfunctions of
${\tilde H}_{n,t_n}$ associated with the eigenvalue $E_n$. There is a subsequence
$(f_{n_j}) \subset (f_n)$ and a function $f \in\mathsf{H}^1(S)$ such that
$f_{n_j} \to f$ weakly in $\mathsf{H}^1(S)$ and weakly in $\mathsf{L}_2(S)$; furthermore,
 we may assume that $f_{n_j} \to f$ in $\mathsf{L}_{2,\rm{loc}}(S)$ and that
$f_{n_j} \to f$ pointwise a.e. All these properties are standard. As a
consequence, for all $\phi \in \mathsf{C}_{\text{c}}^\infty(S)$ we have
$\scapro{\nabla f_{n_j}}{\nabla\phi} \to \scapro{\nabla f}{\nabla\phi}$ and
$\scapro{f_{n_j}}{\phi} \to \scapro{f}{\phi}$. In addition, since translation
is continuous in $\mathsf L_1$ and since $V^{(2)}$ is bounded, we have
$V_{t_n} \to V_{\overline t}$ in $\mathsf L_{2,\text{loc}}(S)$ whence
$$
    \scapro{V_{t_{n_j}} \phi}{f_{n_j}} \to \scapro{V_{\overline t} \phi}{f}, \qquad
   j \to \infty.
$$
Now let ${\mathbf h}_t$, ${\mathbf h}_{n,t}$ and ${\tilde {\mathbf h}}_{n,t}$ denote
the quadratic forms associated with the operators $H_t$, $H_{n,t}$ and
$\tilde H_{n,t}$, respectively, for $t \ge 0$. We note that, for any $g \in {\mathsf H}^1(S)$,
$\phi \in \mathsf C_{\text{c}}^\infty(S)$, and $t \ge 0$ we have
 ${\mathbf h}_{n,t}[g,\phi] = \tilde{\mathbf h}_{n,t}[g,\phi]$  for all sufficiently large $n$.
 Using this and the above convergence properties we may conclude that
$$
  {\mathbf h}_{\overline t}[f,\phi] - E \scapro{f}{\phi}
    = \lim_{j\to\infty} ( {\tilde{\mathbf h}}_{n_j,t_{n_j}}[f_{n_j},\phi] - E
    \scapro{f_{n_j}}{\phi} )
    = 0.
$$
Therefore, $f$ belongs to the domain of $H_{\overline t}$ and satisfies
$H_{\overline t}f = E f$.
} 

It remains to show that ${f} \ne 0$. This follows from the fact that the
eigenfunctions $f_n$ are localized near the interface, i.e., near $\{0\}
\times \S'$.
To prove this, we consider $m \in \N$, $m \le n$, where we observe that
$\psi_m^+ f_n \in D(H_n^+)$ and that
$$
    \lnorm{(\tilde{H}_n^+ - E_n) (\psi_m^+ f_n)} \le 2 \lnorm{\nabla\psi_m^+
    \nabla f_n} + \lnorm{(\Delta\psi_m^+) f_n} \le \frac{c}{m},
$$
where we have also used $[\mathcal{P}_{n,t}, \psi_m^+] = 0$. As a consequence of
\eqref{estimate_alpha} it follows that
$$
   \lnorm{\psi_m^+ f_n} \le c/(\beta m), \qquad n \ge m,
$$
for some constant $c \ge 0$.
A similar estimate works near $-n-t$. We may thus pick $m = m_0$ so large that
$\lnorm{\psi_{m_0}^\pm f_n} \le 1/4$ for all $n \ge m_0$,
whence  $\lnorm{\phi_{m_0} f} \ge 1/2$.
\endproof
%
%
We are now ready for the proof of Thm.~\ref{ThmI}.
\proof[Proof of Theorem~\ref{ThmI}.]
If $E \in \sigma(H_0)$ let $\tau_1 := 0$.
Else Prop.~\ref{solution_of_approximating_problems}
and Lemma~\ref{lem_3.7}  directly yield
a $\tau_1 \ge 0$ such that $E \in \sigma(H_{\tau_1})$; in this case we would
in fact know that $\tau_1 > 0$.

If $E$ happens to be an eigenvalue
of $H_{\tau_1 + 1}$, we let $\tau_2 := \tau_1 + 1$. Else we replace
$V^{(2)}$  with  $V^{(2)} \circ T_{\tau_1 + 1}$,
to obtain some $\tau_2 \ge \tau_1 + 1$ with $E \in \sigma(H_{\tau_2})$, and so on.   \endproof

\section{Examples and Applications}\label{sec_exmp}
For simplicity of notation we restrict our attention in this section mostly
 to the case where $V^{(1)} = V^{(2)}$; we may then simply write $V$.
Also note that, in this section, we do not necessarily require $V \ge 0$.

We first give two examples of potentials $V = V(x,y)$, depending on the
two variables $(x,y) \in \R \times \S'$,  with a (non-trivial) spectral gap
in the spectrum of the associated Hamiltonian.
\exmp In a particularly simple example, $V$ is given as the sum of an
almost periodic (or periodic) potential $V_1 = V_1(x)$ and a potential $V_2 = V_2(y)$,
$$
         V(x,y) := V_1(x) + V_2(y), \qquad (x,y) \in S;
$$
both $V_1$ and $V_2$ are bounded, measurable, and real-valued functions.
Without restriction of generality, we may assume that the spectrum of the
one-dimensional Schr\"odinger operators $h_1 : = - \frac{\d^2}{ \d x^2 } +
V_1(x)$, acting in $\mathsf{L}_2(\R)$, and $h_2 := - \frac{\d^2}{\d y^2} + V_2(y)$,
acting in $\mathsf{L}_2(\S')$, begins at the point $0$.

In addition, let us assume that $h_1$ has a gap $(a,b)$ in its spectrum,
where $0 \le a < b$. In view of condition~\eqref{cond_0.2} let us also
assume that the operator $\ell_{(0,\infty)} + V_1 \!\restriction\! (0,\infty)$
has some essential spectrum in $(-\infty, a]$; here $\ell_{(0,\infty)}$ denotes the
self-adjoint realization of $-\frac{\d^2}{\d x^2}$ in $\mathsf{L}_2(0,\infty)$
with Dirichlet boundary condition at $0$. One may think of
potentials $V_1$ of the type
$V_1(x) := \cos x + \eps \cos \pi x$, with $\eps > 0$ small.
More generally, there are many results that establish the existence of gaps for
certain classes of almost periodic potentials in one dimension (cf.\ e.g.\
Johnson and Moser \cite{JM82}, Avila et al.~\cite{ABD12}, Goldstein and Schlag \cite{GS11}, Puig
\cite{P04}), many of them dealing with the lattice case.

By compactness, $h_2$ has purely discrete spectrum consisting of a sequence of
real eigenvalues $(\la_k)_{k \in \N_0}$ with $\la_0 = 0$, $\la_k < \la_{k+1}$
for all $k \in \N_0$, and $\la_k \to \infty$ as $k \to \infty$.
The condition $\la_0 = 0$ is equivalent to $\int_0^1 V_2(y)\d y = 0$.

The spectrum of the associated Hamiltonian $H_0 = L + V
= h_1 \otimes I + I \otimes h_2$ on $S$ is then given by the algebraic sum
$\sigma(h_1) + \sigma(h_2)$.  In particular,
$H_0$ has a gap in the essential spectrum if $a + \lambda_1 < b$ or
if $\lambda_1 > a$. {\red It is clear that condition~\eqref{cond_0.2} is
satisfied as well.}
We may now apply Thm.~\ref{ThmI} and find that for any given $E > a$ in a gap of $H_0$
there exists a sequence $\tau_k \to \infty$ such that $E$ is an eigenvalue of $H_{\tau_k}$.
\endexmp
\exmp
Another natural class of examples comes with potentials of the form
$$
    V(x,y) = V_1(x,y) + V_2(x,y)
$$
where $V_1$ has a gap in the essential spectrum and $V_2$ decreases at
both ends of $S$. Then (multiplication by) $V_2$ is a relatively compact
perturbation of $L + V_1$ which preserves the essential spectrum. If, in addition,
condition~\eqref{cond_0.2} is met by $V_1$, we can again apply Thm.~\ref{ThmI}.
\endexmp
\exmp\label{exmpThmII}
We now give an application of Thm.~\ref{ThmI} to a dislocation problem on the
plane $\R^2$. 
Suppose $V \colon \R^2 \to \R$ is a bounded, measurable function which
is periodic with period $1$ in the $y$-variable. Let $L$ denote the
(unique) self-adjoint extension of $-\Delta$ defined on
$\mathsf C_{\text{c}}^\infty(\R^2)$ and let
$H = L + V$. It is well known (\cite{DaS78}) that $H$ has
a direct integral decomposition of the form
$$
  H = \int^\oplus_{[0,2\pi)} H(\theta) \frac{\dw \theta }{ 2\pi} ,
$$
where $H(\theta)$ denotes the self-adjoint realization of $-\Delta + V$
in $\mathsf L_2(\R \times (0,1))$ with $\theta$-periodic boundary conditions in the $y$-variable,
 for $\theta \in \R$. The cases $\theta \in 2\pi \Z$ correspond
 to  periodic boundary conditions.
 The spectrum of $\sigma(H(\theta))$ depends continuously on $\theta$
in $[0,2\pi]$ in the usual way and it follows that the spectrum of $H$ has
a band-gap structure. In particular, $H$ has no discrete spectrum.
{\red Furthermore,  the singular continuous spectrum of $H$ is empty provided
the fiber operators $H(\theta)$ have empty singular continuous spectrum
for all $\theta \in [0,2\pi]$ (\cite{DaS78}).}

Let us assume now, in addition, that $H$ has a (non-trivial) spectral gap
$(a,b)$, so that, in particular, $\sigma(H(0)) \cap (a,b) = \emptyset$,
and that $V$ meets condition~\eqref{cond_0.2}.
We obtain $V_t$  from the potential $V$ by  \eqref{dislocpot}
 (with $V^{(1)} =  V^{(2)} = V$ and $(x,y) \in \R^2$), and we let
$H_t(0)$ denote $H(0) + V_t$. By Thm.~\ref{ThmI},
 for any $E \in (a,b)$ there is a sequence of parameters $\tau_k > 0$ such that
$E$ is a (discrete) eigenvalue of $H(0) + V_{\tau_k}$. It is our aim to show that
$H_{\tau_k} = L + V_{\tau_k}$ has a non-zero contribution to the \textit{surface i.d.s.},
the \textit{surface integrated density of states}, in any interval
$(\alpha,\beta) \subset \R$ containing $E$; recall from \cite{EKSchrS, KS01}
that the existence of a surface i.d.s.\ for $H_t$ inside a spectral gap $(a,b)$
of $H_0$ is equivalent to the existence of the limits
\beq\label{surface_ids}
    \mu_{\text{surf}}((\alpha, \beta)) = \lim_{n \to \infty} \frac{1 }{ 2n}
    \tr \E_{(\alpha,\beta)} (H_t^{(n)})
\eeq
for all $a < \alpha < \beta < b$,
where $H_t^{(n)}$ denotes the operator $-\Delta + V_t$ acting in $\mathsf{L}_2(Q_n)$
with periodic boundary conditions, with $Q_n := (-n,n)^2$. For the sake of
comparison, let us also recall that the corresponding formula for the
{\it bulk i.d.s.}\ comes with a normalization factor $1/(4n^2)$.

Under the assumptions made above the limits in  \eqref{surface_ids} may or may not exist,
but there are upper and lower bounds for $ \frac{1}{2n}
    \tr \E_{(\alpha,\beta)} (H_t^{(n)})$:
  First, by Thm.~3.2 of \cite{HK2} we have an upper bound of the form
$$
   \limsup_{n \to \infty} \frac{1 }{ n \log n}  \tr \E_{(\alpha,\beta)}
   (H_t^{(n)}) < \infty,
$$
provided $[\alpha,\beta]$ is contained in a spectral gap of $H_0$;
note that the relevant result in \cite{HK2} does not require any periodicity of $V$.
As for a lower bound, Thm.~\ref{ThmI} implies the following result:
\thm\label{ThmII}
 Let $V =  V(x,y) \colon \R^2 \to \R$ be periodic in $y$ with period $1$
and suppose that $H = -\Delta + V$ has a (non-trivial) spectral gap $(a,b)$. Let us
furthermore assume that
$$
      \min \sigmaess(L_{(0,\infty)} + V \!\restriction\!S^+) \le a,
$$
where $S^+ := (0,\infty) \times (0,1)$.

Then for any interval $\emptyset \ne (\alpha,\beta) \subset [a,b]$  there is a sequence $\tau_k \to
\infty$ such that
%
$$
   \liminf_{n \to \infty} \frac{1 }{ n}  \tr \E_{(\alpha,\beta)}
   (H_{\tau_k}^{(n)}) > 0, \qquad k \in \N.
$$
\endthm\rm
Thm.~\ref{ThmII} generalizes \cite[Cor.~4.5]{HK1} in the sense that periodicity
in $x$ is no longer required. Dropping also the assumption of periodicity in $y$
appears to be much harder.
\proof[Proof of Thm.~\ref{ThmII}.] The proof closely follows the line of argument
in the proof of Thm.~4.3 and Cor.~4.5 of {\red\cite{HK1}:
For any $E \in (\alpha, \beta)$ given,
Thm.~\ref{ThmI} implies that there is a sequence $(\tau_k) \subset \R$ such that
$\tau_k \to \infty$, as $k \to \infty$, and such that $E$ is an eigenvalue of
$H(0) + V_{\tau_k} \! \restriction \! S$. For $k \in \N$, let
 $u_{E,\tau_k} \in D(H(0))$ denote an associated eigenfunction.
 Extending $u_{E,\tau_k}$ by periodicity to all of $\R^2$ we obtain functions
$v_{E,\tau_k}$ which we finally multiply with
cut-off functions $\phi_m(x)\phi_m(y)$ as in the proof of Thm.~4.3 of \cite{HK1} to obtain
$$
         \Psi_{E,\tau_k,m}(x,y) := \phi_m(x) \phi_m(y) v_{E, \tau_k} (x,y), \qquad k,m \in \N.
$$
Then the functions $\Psi_{E, \tau_k, m}$ belong to the domain of the self-adjoint Laplacian
$H$ on $\R^2$ and satisfy
$$
    \norm{(H + V_{t_k} - E) \Psi_{E,\tau_k,m}} \le \frac{c }{ m}
    \norm{\Psi_{E,\tau_k,m}}.
$$
We may now continue as in the proof of Cor.~4.5 in \cite{HK1}.}
\endproof

\endexmp

\exmp[Half-space problems]\label{half-space-problems}
Suppose $V^{(2)}$ satisfies the conditions of Theorem~\ref{ThmI} and let
$V^{(1)} = 0$.  Then $H_t$ is a model
for a semi-infinite nano-tube while the corresponding $y$-periodic
extension to all of $\R^2$ yields a model for a half-crystal, i.e.,
for a facet of a crystal. {\red Consider some $E$ inside a gap $(a_0,b_0)$.
Then Thm.~\ref{ThmI} states that, for suitable $t = \tau_k$,
we have a bound state of $H_{\tau_k}$ at $E$  for the
half-infinite tube. Furthermore, in the case of a half-crystal,
we will have a non-trivial i.d.s.\ for $t =\tau_k$ near $E$
 by Thm.~\ref{ThmII}.}
\endexmp
\section{Lipschitz continuity of discrete eigenvalues}\label{sec_lipschitz}
In this section, we establish continuity of the discrete eigenvalues of
dislocation problems and, under an additional assumption on the potential,
also (local) Lipschitz continuity, as stated in Thm.~\ref{thm_lipschitz}.

As a preparation for the proof of Thm.~\ref{thm_lipschitz}
let us first give a detailed account of the coordinate transformation used in
Section~\ref{sec_prelim}. 
  It will be convenient to reverse the
direction of dislocation in order to avoid a host of irritating minus-signs.
Furthermore, we restrict our attention to
the case where $V^{(1)} = V^{(2)}$ which allows us to write $V$; the proof in
the general case is virtually the same. Hence the dislocation potential $V_t$ for
$t \in \R$ is now given by $V_t(x,y) = V(x,y)$, for $x < 0$,
and by $V_t(x,y) = V(x-t, y)$, for $x > 0$.
For simplicity, we assume again $V \ge 0$.
It is our aim to compare the quadratic forms for the dislocation problem
in the coordinates $(x,y)$ with the
corresponding forms in suitably transformed variables.

We will work in the following well-known setting:
 Let  $\mathcal H_1$, $\mathcal H_2$ be Hilbert spaces
and let ${\bf a}_1$ and ${\bf a}_2$ be (densely defined)
sesquilinear forms with domains  $D({\bf a}_1) \subset{\mathcal H}_1$ and
 $D({\bf a}_2) \subset {\mathcal H}_2$.
We say that the forms ${\bf a}_1$ and ${\bf a}_2$ are
{\it unitarily equivalent} if there exists a unitary map
$U \colon  {\mathcal H}_1 \to {\mathcal H}_2$ such that
$D({\bf a}_2) = U D({\bf a}_1)$ and ${\bf a}_2[u,v] = {\bf a}_1[U^{-1}u,
U^{-1} v ]$, for all $u, v \in D({\bf a}_2)$.
 If, in addition, the forms ${\bf a}_1$, ${\bf a}_2$ are
semi-bounded and closed, then the associated self-adjoint operators $H_1$ and
$H_2$ are unitarily equivalent:
$H_2 = U H_1 U^{-1}$ and $D(H_2) = UD(H_1)$. This is immediate from
the first representation theorem for quadratic forms.

We will need a family of simple diffeomorphisms of the real line.
Here a particularly efficient construction is to start
from piecewise linear (continuous) functions and then to apply
a Friedrichs mollifier; this gives sufficient control of the function
and its derivatives. For $t \in [-1/2,1/2]$, let ${\tilde \varphi}_t \colon {\R}
\to {\R}$ be defined by ${\tilde \varphi}_t(x) = x$ for
$x \le 0$, $\tilde \phi_t(x) = (1 + t)x$ for $0 \le x \le 1$,
and  ${\tilde \varphi}_t(x) = x + t$ for $x \ge 1$. We now let
%
$$
     \varphi_t := j_1 * {\tilde \varphi}_t, \qquad |t| \le 1/2,
$$
where $j_\eta$, for $\eta > 0$, is the usual kernel of the Friedrichs mollifier.
Also note that $\phi_{0} =\text{id}$.  There exists a constant $c \ge 0$ such that
%
\beq\label{properties_of_phi}
   |\varphi_{t}(x) - x| \le c |t|, \quad
   |\varphi'_{t}(x) - 1| \le |t|, \quad
   |\varphi_{t}''(x)| \le c |t|,
\eeq
for $x \in \R$ and $|t| \le 1/2$.
The first and second estimate follow directly from standard properties of
the Friedrichs mollifier and the definition of ${\tilde \varphi}_t$. As for
the last inequality, we note that the second order distributional derivative
of ${\tilde \varphi_t}$ is given by $t \delta_0 - t \delta_1$
with $\delta_0$ and $\delta_1$ denoting
the Dirac distributions at the points $0$ and $1$, respectively, and then
$\varphi_{t}'' = t j_1 - t j_1(\cdot - 1)$. For the remainder of this section
we will denote by $c$ various non-negative constants with the understanding
that the value of $c$ may change from one line to the next.

We define diffeomorphisms $\Phi_t \colon {S} \to {S}$  by setting
$$
   (\xi, \eta)^T = \Phi_t(x, y) = (\varphi_t(x), y)^T;
$$
notice that $\eta = y$.
The determinant of the Jacobi matrix of $\Phi_t(x,y)$ is $\varphi_t'(x)$,
with $1/2 \le \varphi_t'(x) \le 3/2$,  for $|t| \le 1/2$ and $x \in \R$.
Then the mapping
$$
  U_t \colon\mathsf L_2({S};\dw \xi \dw\eta) \to \mathsf L_2({S}; \varphi'_t(x)\, \dw x \dw y),
  \qquad
  u \mapsto v = U_t u = u \circ \Phi_t,
$$
is unitary; furthermore, it is clear that
$U_t {\mathsf H}^1({S}) =  {\mathsf H}^1({S})$.
Routine calculations show that, for $|t| \le 1/2$,  the form
$$
    {\mathbf a}_t[u] := \int_{S} |\nabla u(\xi,\eta)|^2\dw \xi \dw\eta +  \int_{S} V_t(\xi,\eta)
    |u(\xi,\eta)|^2\dw \xi \dw\eta,
 $$
defined on $D({\mathbf a}_t) :=  {\mathsf H}^1({S})$
in the Hilbert space $\mathsf L_2({S};\dw\xi\dw\eta)$, is unitarily equivalent to the form
$$
    {\mathbf b}_t[v] :=  \int_{S}  \left( \frac{1 }{ \varphi_t'} |\partial_1 v|^2 + \varphi_t'
   |\partial_2 v|^2\right) \dw x \dw y
    +  \int_{S} V_t(\Phi_t(x,y)) |v(x,y)|^2 \varphi_t'\,\dw x \dw y
$$
on $ D({\mathbf b}_t) =  {\mathsf H}^1({S})$ in the Hilbert space
 $\mathsf L_2({S}; \varphi_t'\,\dw x \dw y)$.
The mapping
$$
   \tilde U_t \colon\mathsf L_2({S}; \varphi'_t\,\dw x \dw y) \to\mathsf L_2({S}; \dw x \dw y),
   \qquad
  v \mapsto w = \sqrt{\varphi_t'} v,
$$
is unitary and  $\tilde U_t {\mathsf H}^1({S}) =  {\mathsf H}^1({S})$.
From $v = \frac{1 }{ \sqrt{\varphi_t'}} w$ we get by
straight-forward calculations that the form ${\bf b}_t$ is unitarily equivalent to the form
\begin{align}
  {\bf c}_t[w] & :=
 \int_{S} \left( \frac{1 }{ (\varphi_t')^2} |\partial_1 w |^2
     + |\partial_2 w|^2
     -  \frac{\varphi_t'' }{ (\varphi_t')^3}\, \hbox{\rm Re}\, ({\bar w}  \partial_1 w )
     +    \frac{(\varphi_t'')^2 }{ 4  (\varphi_t')^4} |w|^2
             \right)\dw x \dw y \cr
     & \qquad \qquad \qquad +  \int_{S} V_t(\Phi_t(x,y)) |w|^2\dw x \dw y,
\nonumber
\end{align}
in $\mathsf L_2({S})$, again with domain $D({\bf c}_t) = {\mathsf H}^1({S})$.

We wish to apply the perturbation theorem for quadratic forms \cite[Thm.~VI-3.9]{K66}
to the family $({\bf c}_t)_{|t| \le 1/2}$ with ${\bf c}_0$
playing the r\^ole of the unperturbed form. This means that we take
\begin{align}
   {\bf d}_t[w]
       & := {\bf c}_t[w] - {\bf c}_0[w] \cr
       & =   \int_{S}  \left(\left(\frac{1 }{ (\varphi_t')^2} - 1\right) |\partial_1 w|^2
                     - \frac{\varphi_t'' }{ (\varphi_t')^3}\, \hbox{\rm Re}\, ({\bar w}
                                        \partial_1 w )
        +  \frac{(\varphi_t'')^2 }{ 4  (\varphi_t')^4} |w|^2\right)\dw x\dw y \cr
        & \qquad \qquad +   \int_{S} \left(V_t(\Phi_t(x,y)) - V(x,y)\right) |w|^2\dw x\dw y
\label{d_t}
\end{align}
with domain $\mathsf{H}^1(S)$ as a perturbation and check that the assumptions of
\cite[Thm.~VI-3.9]{K66} are satisfied.
Define $C_t$ as the self-adjoint operator associated with the
quadratic form ${\bf c}_t$, for $|t| \leq 1/2$.
%
We have the following lemma.
\lem\label{lem_nrc} Let $0 \le V \in \mathsf L_\infty(S)$ and let $C_t$,
for $|t| \le 1/2$, be as above. We then have:
\vskip.5ex
 $(a)$ $(C_t + 1)^{-1} - (C_0 + 1)^{-1}$ is compact for
all $|t| \leq 1/2$.

\vskip.5ex
 $(b)$ Let $\zeta \in \varrho(C_0)$. Then there exists $\tau_0 = \tau_0(\zeta) > 0$
such that $\zeta \in \varrho(C_t)$ for $|t| \le \tau_0$ and
$\norm{(C_t - \zeta)^{-1} - (C_0 - \zeta)^{-1}} \to 0$ as $t \to 0$.

\vskip.5ex

 $(c)$ Assume, in addition, that $\partial_1 V$ is a signed Borel measure,
and let $\zeta \in \varrho(C_0)$.  Then there exists $\tau_1 = \tau_1(\zeta) > 0$
such that $\zeta \in \varrho(C_t)$ for $|t| \le \tau_1$ and
\beq\label{norm_ct-c0}
   \norm{(C_t - \zeta)^{-1} - (C_0 - \zeta)^{-1}} \le c |t|,
    \qquad 0 \le |t| \le \tau_1.
\eeq
\endlem\rm
\vskip.5ex

The first statement of the lemma implies $\sigmaess(C_t) = \sigmaess(C_0)$ which
in turn implies $\sigmaess(H_t) = \sigmaess(H_0)$ by the unitary equivalence of
$C_t$ and $H_t$.

\proof
(a) By the second resolvent equation and  \eqref{d_t}, we have to study
several terms like
$$
  I_t := (C_t + 1)^{-1} \partial_1 (1 - 1/(\phi_t')^2) \partial_1
 (C_0 - 1)^{-1},
$$
and also the term
$$
  J_t := (C_t + 1)^{-1} (V_t \circ \Phi_t - V) (C_0 + 1)^{-1}.
$$
Notice that the functions $1 - 1/\phi_t'(x)^2$ and $V_t \circ \Phi_t - V$ have
support in $K := [-2, 3] \times \S'$ and are (uniformly) bounded.
For all these terms compactness is immediate by the Rellich Compactness Theorem.
\vskip.5ex
(b) With the aid of \eqref{properties_of_phi}
it is easy to see that the first three terms on the right hand side of \eqref{d_t}
are bounded by $ c |t|  (\mathbf c_0[w] + \norm{w}^2)$, for $w \in
\mathsf{H}^1(S)$, where $c \ge 0$ is a
suitable constant. We next provide an estimate for
$ \int_S | V_t \circ \Phi_t - V| |w|^2 \dw x \dw y$; let us
 write $Y_t := |V_t \circ \Phi_t - V|$ for short.

From a simple Sobolev estimate (which we will prove below in part (d)) one obtains that there is
a constant $c \ge 0$ such that
$\norm{\chi_K (C_0 + 1)^{-1/2} u}_{\mathsf L_3(S)} \le c \norm{u}$ for $u \in \mathsf L_2(S)$,
whence
\begin{align}
    \scapro{Y_t (C_0 + 1)^{-1/2}u}{ (C_0 + 1)^{-1/2}u}
  &  =  \int_S  Y_t |\chi_K (C_0 + 1)^{-1/2}u|^2 \d x \d y \nonumber \\
  &  \hskip-5em \le \norm{Y_t}_{\mathsf L_3(S)} \cdot \norm{\chi_K (C_0 + 1)^{-1/2}u}_{\mathsf L_3(S)}^2
   \le c \norm{Y_t}_3 \cdot \norm{u}^2. \nonumber
\end{align}
We have thus shown that
%
$$
   \int_S |V_t \circ \Phi_t - V| |w|^2 \d x \d y \le c  \norm{ V_t \circ \Phi_t - V}_{\mathsf L_3(S)}
                     (\mathbf c_0[w] + \norm{w}^2), \qquad w \in D(\mathbf c_0).
$$
Combining what we have obtained so far and writing $\gamma(t) :=  \norm{ V_t \circ \Phi_t - V}_3$,
 we now see that
\beq\label{form_estimate_d_t}
  |\mathbf d_t[w]| \le c(|t| + \gamma(t))(\mathbf c_0[w] + \norm{w}^2),
   \qquad   w \in D(\mathbf c_0).
\eeq
Here dominated convergence yields $ \gamma(t) \to 0$,
as $t \to 0$, and the statement $(b)$ now follows from \cite[Thm.~VI-3.9]{K66}.
Note that both of the constants $a$ and $b$ in \cite[loc.~cit.]{K66} correspond to the same factor
$c (|t| + \gamma(t))$ in  \eqref{form_estimate_d_t} and that
$\norm{C_0 (C_0 - \zeta)^{-1}} \le 1 + \gamma_0 |\zeta|$ with $\gamma_0
:= \norm{(C_0 - \zeta)^{-1}}$, whence
$$\norm{(a + b C_0) (C_0 - \zeta)^{-1}} \le c(|t| + \gamma(t)) (1 + |\zeta|)\gamma_0.$$
\vskip.5ex

(c) Suppose in addition that $\partial_1 V$ is a signed measure.
Let $\zeta \in \mathsf C_{\mathsf c}^\infty(\R)$ satisfy $\zeta(x) = 1$ for $-2 \le x \le 3$
and write $\tilde V := \zeta V$.
We then have ${V}_t \circ \Phi_t - {V} = {\tilde V}_t \circ \Phi_t - {\tilde V}$
for $|t| \le 1/2$. Applying Lemma~\ref{JVlem2} with $W := \tilde V$ we find that
\beq\label{estimate_JVlem2}
    \norm{({V}_t \circ \Phi_t - {V})}_1 \le c |t|, \qquad |t| \le t_0.
\eeq
Since $V$ is bounded and $\supp (V_t \circ \Phi_t - V) \subset K$, this implies
$\norm{ V_t \circ \Phi_t - V}_3 \le c |t|$, for $|t| \le 1/2$.
Inserting this in  \eqref{form_estimate_d_t}, we obtain
$$
   |\mathbf d_t[u]| \le c t (\mathbf c_0[u] + \norm{u}^2),
   \qquad   u \in D(\mathbf c_0),
$$
and the desired result follows as in part (b) from \cite[Thm.~VI-3.9]{K66}.

(d) We finally give the details of the Sobolev estimate used in part (b).
Let $B$ denote a disc of radius $1/2$ in $S$ and let
$\psi \in  \mathsf{C}_{\text c}^\infty(B)$. For any $v \in \mathsf H^1(S)$,
we have $\psi v \in \mathsf W^1_p(B)$, for any $1 \le p \le 2$; furthermore,
there are constants $c_p \ge 0$ such that
$$
    \norm{\psi v}_{\mathsf W^1_p(B)} \le c_p \norm{v}_{\mathsf H^1(B)}, \qquad v \in \mathsf H^1(S).
$$
The specific choice $p := 6/5$ and the Gagliardo-Nirenberg-Sobolev inequality
(\cite{Ev98}) yields
an estimate $\norm{\psi v}_{\mathsf L_3(B)} \le c \norm{v}_{\mathsf H^1(B)}$, for all
$v \in \mathsf H^1(S)$.
 Using a covering of $K$ by a finite number of discs $B_i$ of radius $1/2$
 and a suitable partition of unity $\{\psi_i\}$ subordinate to this covering,
 we easily obtain an estimate of the form
$$
     \norm{v}_{\mathsf L_3(K)} \le C \norm{v}_{\mathsf H^1(S)}, \qquad v \in \mathsf H^1(S),
$$
with a constant $C \ge 0$.
\endproof

Before passing to the proof of Thm.~\ref{thm_lipschitz} let us give some details on
what the notion of continuity of eigenvalues is supposed to mean (cf.\ also
the appendix in {\red \cite{HK1}}). Suppose $a_0 < b_0 \in \R$ are such that
$a_0, b_0 \in \sigmaess(H_0)$ and $(a_0, b_0) \cap \sigmaess(H_0) = \emptyset$.
Then the discrete eigenvalues of $H_t$ in $(a_0,b_0)$ can be described by a
countable family $(f_i)$ of continuous functions $f_i \colon (\alpha_i,\beta_i) \to
(a_0, b_0)$ (where $\alpha_i < \beta_i$ and $\alpha_i \in \R \cup \{-\infty\}$,
$\beta_i \in \R \cup \{+\infty\}$) such that any compact set
$K_{T, a, b} := [-T,T] \times [a,b] \subset \R \times (a_0,b_0)$ meets only finitely many of
the graphs $\Gamma(f_i)$.  Notice that we may have to deal with multiple eigenvalues.
We may agree that each of the functions $f_i$ carries precisely spectral multiplicity 1,
so some of the functions might be identical, or their graphs may overlap or intersect.
We do not discuss continuity of the associated eigenprojections here. Also note
that, in general, the description of the eigenvalues by the family of functions
$(f_i)$ will not be unique.

More precisely, if $E \in (a_0,b_0)$ and $t \in \R$
are such that $E \in \sigma(H_t)$, then there exists some $i \in \N$ with
$f_i(t) = E$. Conversely, if $\alpha_j < t < \beta_j$ for some $j$,
then $f_j(t)$ is an eigenvalue of $H_t$. In addition, the graphs $\Gamma(f_i)$ \textit{leave
any of the sets $K_{T,a,b}$ as $t \downarrow \alpha_i$ and $t \uparrow \beta_i$} in the
following sense: if, e.g., $-T \le \alpha_i < T$ for some $i$, then there
exists some $\alpha_i' \in (\alpha_i, T)$ such that $f_i(t) \notin [a,b]$
for $\alpha_i < t < \alpha_i'$ and $f_i(\alpha_i') = a$ or $f_i(\alpha_i') = b$.
Similar statements hold for the case where $\alpha_i < -T$ and for $\beta_i
\in (-T,T]$ or $\beta_i > T$. It follows, in particular, that $f_i(t) \to a_0$ or
$f_i(t) \to b_0$ as $t \downarrow \alpha_i$, if $\alpha_i$ is finite, and similarly
for $t \uparrow \beta_i$.

\proof[Proof of Thm.~\ref{thm_lipschitz}]  Since $H_t$ and $C_t$ are unitarily equivalent
for $|t| \le 1/2$, it is enough to prove the corresponding statements for
the operators $C_t$. It is obvious that this implies the statements for all $t \in \R$
since, in view of Remark~\ref{remark_Vt_Lipschitz}, we may replace $V$ with any $V_t$;
indeed, if $\partial_1 V$ is (locally) a signed measure, the same is true for
$V_t$.

(a) By the preceding Lemma~\ref{lem_nrc}, the resolvents $(C_t + 1)^{-1}$ depend continuously
in norm on $t$. Suppose $E_0$ is a discrete eigenvalue of $C_{t_0}$ for some $t_0 \in
-[1/4, 1/4]$. Then there are $\eps_0 > 0$ and $\tau_0 > 0$
such that $E_0$ is the only eigenvalue of $C_{t_0}$ in  $(E_0 - 3\eps_0, E_0 + 3\eps_0)$
 and such that $C_t$ has no spectrum in $(E_0 - 3\eps_0, E_0-\eps_0) \cup (E_0 +\eps_0,
E_0 + 3\eps_0)$ for $|t - t_0| \le \tau_0$. It follows by well-known results
\cite[Thm.~VIII-20]{RS-I} that $\tr \E_{(E_0 - 2\eps_0, E_0 + 2\eps_0)}$ is constant for $|t - t_0| \le \tau_0$.
Consider $\zeta' := E_0 - 2\eps_0$.
 Then min-max (\cite[Exercise XIII.2]{RS-IV}) implies that the maximum value of the
 spectrum of $(C_t - \zeta')^{-1}$ is a continuous eigenvalue branch, for
$|t - t_0| \le \tau_0$. This branch corresponds to a continuous eigenvalue branch
of $C_t$ which passes through the point $(t_0,E_0)$ and has the largest value among
all branches passing through $(t_0,E_0)$. If the eigenvalue $E_0$ has multiplicity
greater than $1$, the above method will yield {\red a corresponding number of
 continuous eigenvalue branches below (or equal) to the first branch.}

 This local description of the eigenvalues of $H_t$ in terms of continuous functions
can be patched together by means of simple compactness arguments to yield a ``global'' picture;
cf.\ also the appendix in \cite{HK1}. Indeed, if $a_n \downarrow a_0$, $b_n \uparrow b_0$, and
$T_n \to \infty$ monotonically (with $a_0 < a_n < b_n < b_0$), we let
$K_n := K_{T_n,a_n,b_n}$ so that $K_n \subset K_{n+1}$ and $\cup_n K_n = \R \times (a_0,b_0)$.
Then for each pair $(t_0,E_0) \in K_n$ with the property that $E_0 \in \sigma(C_{t_0})$ we have an open
neighborhood
$$
   U_{t_0,E_0;\tau_0,\eps_0} := (t_0-\tau_0, t_0+ \tau_0) \times (E_0 - 3\eps_0, E_0 + 3\eps_0)
$$
with the properties obtained above; here $\tau_0$ and $\eps_0$ are positive numbers.

\noindent If $(t_0,E_0)$ is such that $E_0 \notin \sigma(C_{t_0})$ there
is an open neighborhood $\tilde U_{t_0,E_0;\tau_0,\eps_0}$ with the property that $C_t$ has no spectrum
in $(E_0 - \eps_0, E_0 + \eps_0)$ for $|t| \le \tau_0$.
By compactness, a finite selection of the sets
$U_{t_0,E_0;\tau_0,\eps_0}$ and $\tilde U_{t_0,E_0;\tau_0,\eps_0}$ covers $K_n$, etc.

(b) We will show that the functions $f_i \colon (\alpha_i,\beta_i) \to \R$ introduced above are
locally (uniformly) Lipschitz, i.e.,
for any compact interval $[\tilde\alpha, \tilde\beta] \subset (\alpha_i,\beta_i)$ there
exists a constant $c\ge 0$ such that $|f_i(t) - f_i(t')| \le c |t - t'|$, for all
$t, t' \in [\tilde\alpha, \tilde\beta]$. We continue in the same setting as in part (a) of this proof.
Let $E_0$ be a discrete eigenvalue of $C_{t_0}$ for some $|t_0| < 1/4$, and let
$\tau_0 > 0$, $\eps_0 > 0$, and $\zeta'$ as above. Replacing $V$ with $V_t$ in  \eqref{norm_ct-c0}
(and making also use of Remark \ref{remark_Vt_Lipschitz}) we obtain an estimate
analogous to  \eqref{norm_ct-c0} for the pair $(t,t')$ in place of the pair
$(0,t)$, viz.,
$$
  \left| \mathbf c_t[u] - \mathbf c_{t'}[u] \right| \le c |t - t'| (\mathbf c_t[u] + \norm{u}^2),
  \qquad u \in D(\mathbf c_0),
$$
with a constant $c \ge 0$ which can be chosen independently of $t, t' \in
 (t_0 - \tau_0, t_0 + \tau_0)$.
Evoking again \cite[Thm.~VI-3.9]{K66} we obtain a resolvent estimate
$$
   \norm{(C_t - \zeta')^{-1} - (C_{t'} - \zeta')^{-1}} \le \frac{c}{\eps_0^2} |t - t'|, \qquad
   t, t' \in (t_0 - \tau_0, t_0 + \tau_0),
$$
where $c \ge 0$ is a constant,
and  min-max as in \cite[Exercise XIII.2]{RS-IV} gives the desired result.
\endproof
We finally discuss how to establish the crucial estimate \eqref{estimate_JVlem2}.
\lem\label{JVlem2} Let $\alpha\in \mathsf C^1(\R,\R)$ with $\norm{\alpha}_\infty<\infty$ and $\norm{\alpha'}_\infty\leq 1/2$
be given, and let $\phi\colon\R^2\to\R^2$ be defined by $\phi(x_1,x_2)=(x_1+\alpha(x_1),x_2)$.
Let $W\in\mathsf L_1(\R^2)$, and assume that the distributional derivative $\partial_1 W$ is a (signed)
measure $\mu$ of finite total variation $\norm{\mu}_1$.

Then $\norm{W\circ\phi-W}_1 \leq 2 \norm{\mu}_1\norm{\alpha}_\infty$.
\endlem
\proof (1) In a first step, we show the assertion under the additional hypothesis that
$W\in\mathsf C^1(\R^2)$ and $\partial_1W\in\mathsf L_1(\R^2)$. From
$$\frac{\dw}{\dw s}W(x_1+s\alpha(x_1),x_2) = \partial_1W(x_1+s\alpha(x_1),x_2)\alpha(x_1)$$
we then obtain
\begin{align}
\norm{W\circ\phi-W}_1 & = \int\left|\int_0^1\partial_1 W(x_1+s\alpha(x_1),x_2)\d s\;\alpha(x_1)\right|\d x \nonumber\\
&\leq\norm{\alpha}_\infty\int_0^1\norm{\partial_1 W\circ\phi_s}_1\d s\label{JV1}
\end{align}
with $\phi_s\colon\R^2\to\R^2$, $\phi_s(x_1,x_2)=(x_1+s\alpha(x_1),x_2)$.
From $1+s\alpha'(x_1)\geq 1/2$ for all $s\in[0,1]$, $x_1\in\R$, one obtains
that $\phi_s$ has an inverse $\psi_s$ with the property that $|\det\psi_s'|\leq 2$.
Using the substitution $x=\psi_s(y)$, $\dw x=|\det\psi_s'(y)|\d y$ one obtains
$$\int|\partial_1W(\phi_s(x))|\d x = \int|\partial_1W(y)||\det\psi_s'(y)|\d y \leq 2\norm{\partial_1W}_1 .$$
Inserting this inequality into \eqref{JV1} one obtains the assertion for the present special case.

(2) In order to prove the general case, let $(\rho_k)\subset\mathsf C_{\text{c}}^\infty(\R^2)$ be a $\delta$-sequence. It is then
standard to show that $\rho_k*W\to W$ in $\mathsf L_1(\R^2)$, $\rho_k*W\in\mathsf C^1(\R^2)$ and $\partial_1(\rho_k*W)\in\mathsf L_1(\R^2)$,
where $\partial_1(\rho_k*W)(x)=\int\rho_k(x-y)\d\mu(y)$ and $\norm{\partial_1(\rho_k*W)}_1\leq\norm{\mu}_1$.
Using the case treated in (1) we conclude that
\begin{align}
\norm{(\rho_k*W)\circ\phi - \rho_k*W}_1 & \leq 2\norm{\partial_1(\rho_k*W)}_1\norm{\alpha}_\infty \nonumber\\
& \leq 2\norm{\mu}_1\norm{\alpha}_\infty\nonumber
\end{align}
and for $k\to\infty$ we obtain the assertion.
\endproof
\rem If $W\in\mathsf L_1(S)$ and the distributional derivative $\partial_1 W$ is a measure $\mu$ (on $S$) of finite
total variation, the same inequality as in Lemma~\ref{JVlem2} holds.
\endrem
In fact, our assumptions on the potential $V$ are almost optimal, which follows from the next lemma.
\lem\label{JVlem1} Let $f\in\mathsf L_1(\R^n)$ and $C\geq 0$.

Then the following properties are equivalent:
\begin{enumerate}
\item[(1)] The mapping $\R\to\mathsf L_1(\R^n)$, $t\mapsto f(\cdot -te_1)$ is Lipschitz
continuous with Lipschitz constant $C$.
\item[(2)] The distributional derivative $\partial_1 f$ is a (signed) Borel-measure
of finite total variation $\leq C$.
\end{enumerate}
\endlem\rm
\proof $(1)\Longrightarrow(2):$ Let $\phi\in\mathsf C_{\text{c}}^\infty(\R^n)$. Then
\begin{align}
\partial_1f(\phi) & = - \int f(x)\partial_1\phi(x)\d x \nonumber\\
& = -\lim_{t\to 0}\int f(x)\frac{1}{t}(\phi(x+te_1)-\phi(x))\d x\nonumber\\
& = -\lim_{t\to 0}\frac{1}{t}\int(f(x-te_1)-f(x))\phi(x)\d x.\nonumber
\end{align}
Using $\norm{f(\cdot-te_1)-f}_1\leq Ct$ we obtain
$$|\partial_1f(\phi)|\leq C\norm{\phi}_\infty.$$
Now the Riesz-Markov Theorem implies the assertion.

$(2)\Longrightarrow(1):$ For $t>0$, $\phi\in\mathsf C_{\text{c}}^\infty(\R^n)$ we compute, using Fubini's theorem,
\begin{align}
\int(f(x-te_1)-f(x))\phi(x)\d x & = \int f(x)(\phi(x+te_1)-\phi(x))\d x \nonumber\\
& = \int_0^t\int f(x)\partial_1\phi(x+se_1)\d x\d s \nonumber\\
& = -\int_0^t\partial_1f(\phi(\cdot+se_1))\d s.\nonumber
\end{align}
This implies
$$\left|\int(f(x-te_1)-f(x))\phi(x)\d x\right|\leq t\norm{\partial_1f}_{\text{var}}\norm{\phi}_\infty$$
and therefore
\begin{align}
\norm{f(\cdot-te_1)-f}_1 & = \sup\left\{\left|\int(f(x-te_1)-f(x))\phi(x)\d x\right|;\,\phi\in
\mathsf C_{\text{c}}^\infty(\R^n),\,\norm{\phi}_\infty\leq 1\right\}\nonumber\\
& \leq t\norm{\partial_1f}_{\text{var}}\nonumber\\
& \leq Ct.\nonumber
\end{align}
\endproof
\rem\label{remark_Vt_Lipschitz}
Let $W \in \mathsf L_1(S) \cap\mathsf L_\infty(S)$ satisfy the assumptions on $f$ in Lemma~\ref{JVlem1} and
let ${v} \in \mathsf L_\infty(S)$ satisfy the condition
$$
   \norm{{v}(.-te_1) - {v}}_1 \le c |t|, \qquad t \in \R.
$$
Then the product ${v} W$ enjoys the same properties as $W$. Indeed,
\begin{align}
  \norm{({v} W)(. - te_1) - {v} W}_1 & \le \norm{W}_\infty \norm{{v}(.-te_1) - {v}}_1
 +   \norm{{v}}_\infty \norm{W(.-te_1) - W}_1 \nonumber\\
   & \le C|t|,
\nonumber
\end{align}
and the assertion follows from Lemma \ref{JVlem1}.

On $\S'$, characteristic functions $\chi$ of
the form $\chi = \chi_{(a,\infty) \times \S'}$ satisfy the conditions on $v$ in
this remark. We therefore see, in particular, that $\partial_1 (V_t)$
is (locally) a signed measure (for any $t \in \R$) if $\partial_1 V$ has this property.
\endrem
\section{Appendix: Hilbert-Schmidt properties}\label{appendix}

%
%
%
%
In the following, we prove Lemma~\ref{lem2.1} and Lemma~\ref{lem2.2}.

(1) We start from the free resolvent of $H_0$, the Laplacian in $\R^2$; note that
we now write $x = (x_1,x_2) \in \R^2$ etc. It is
well-known that $(H_0 + 1)^{-1}$ is an integral operator with kernel given
by
$$
    G_0(x,y) :=  (H_0 + 1)^{-1}(x,y) = {\mathbf K}_0(|x-y|), \qquad x, y \in \R^2,
$$
where $ {\mathbf K}_0 \in\mathsf C^\infty(0,\infty)$ is the modified Hankel function of order $0$
(\cite[p.~127]{M88}, \cite[\'Eqn.~VII-10.15, p.~286]{S73}, {\red\cite{O80}}; also
\cite[p.~128, Exercise 49]{RS-II}).
$\mathbf{K}_0$ is a smooth, monotonically decreasing function.
The asymptotic behavior of $\mathbf{K}_0$ for $r \downarrow 0$ and for $r \to \infty$ as
in \cite[Par.~28, Lemma 8 and Thm.~3]{M88} gives an estimate
\beq
 \mathbf{K}_0(r) \le \left\{
 \begin{array}{ll}
   c\, (1 + |\log r|), & 0 < r \le 1, \\
   c\, \ee^{-r}, & r \ge 1,
 \end{array}\right.
 \label{K0}
\eeq
for some constant $c \ge 0$.

(2) Let $S' := \R \times (0,1)$ and let $H_{S'}$ denote the Laplacian
of $S'$ with periodic boundary conditions in the $x_2$-variable.
$H_{S'}$ is non-negative and self-adjoint. The operator $(H_{S'}+1)^{-1}$
has an integral kernel $G_{S'} = G_{S'}(x,y)$ which can be computed in terms of
$\mathbf{K}_0$ by the classical method of image charges 
$$
 G_{S'}(x,y) = \sum_{k \in \Z} G_0(x + k \mathbf{e}_2, y)
       = \sum_{k \in \Z} \mathbf{K}_0(|x + k \mathbf{e}_2 - y|),
   \qquad x, y \in S'.
$$
Convergence of the sum in $(2)$ follows from the exponential decay of $\mathbf{K}_0$.
Also note that $G_{S'} \ge 0$.

(3) We now introduce an additional Dirichlet boundary condition on the
line segment $\ell := \{(x_1,x_2) \in \R^2 ;  x_1 = 0,  0 < x_2 < 1\}$ which
decouples the strip $S'$ into a left part $S'_-$ and a right part $S'_+$.
Let $H_{S'_\pm}$ denote the Laplacian of $S'_\pm$ with Dirichlet boundary condition
on $\ell$ and periodic boundary conditions with respect to the $x_2$-variable.
Let $H_{S',\dec} = H_{S'_-} \oplus H_{S'_+}$, and denote the associated resolvent
kernels by $G_{S'_\pm}$ and  $G_{S',\dec}$, respectively.
 {\red Path integral methods imply that
$$
   0 \le G_{S',\dec}(x,y) \le G_{S'}(x,y),
   \qquad x, y \in S'_+ \text{\ or\ } x, y \in S'_-;
$$
cf.\ \cite{DS76} or  \cite[Thm.~2.1.6]{D89}.
In order to apply the method of image charges,
we write $x^* := (-x_1, x_2)$ for $x = (x_1,x_2) \in S'$.
We then have
$$
   G_{S'_+}(x,y) = G_{S'}(x,y) - G_{S'}(x^*,y), \qquad x, y \in S'_+,
$$
and similarly for $S'_-$, and we see that
$$
   0 \le G_{S'}(x,y) - G_{S'_\pm}(x,y)
    = G_{S'}(x^*,y) = \sum_{k \in \Z}\mathbf{K}_0(|x^* + k \mathbf{e}_2 - y|),
$$
for $x, y \in S'_+$ and for $x, y \in S'_-$.

If $x \in S'_\pm$ and $y \in S'_\mp$ we extend $G_{S',\dec}$ by zero and have
$$
  G_{S'}(x,y) - G_{S',\dec}(x,y) = G_{S'}(x,y) = \sum_{k \in \Z}\mathbf{K}_0(|x + k \mathbf{e}_2 - y|),
$$
for $(x,y) \in S'_+ \times S'_- \cup S'_- \times S'_+$.

(4) Writing $K(x,y) :=  G_{S'}(x,y) - G_{S'_-}(x,y) - G_{S'_+}(x,y)$ we now show that
$K \in\mathsf L_2(S' \times S')$, using the estimate of  \eqref{K0}. Let us first consider the
case where $x, y \in S'_+$. Then
$$
    |x^* + k \mathbf{e}_2 - y| = \sqrt{(x_1 + y_1)^2 + (k + x_2 - y_2)^2}, \qquad k \in \Z,
$$
so that by the monotonicity of $\mathbf{K}_0$,  \eqref{K0}, and $-1 < x_2 - y_2 < 1$,
$$
    K(x,y) \le 3\mathbf{K}_0(\sqrt{(x_1 + y_1)^2}) + 2\sum_{k\in\N} \mathbf{K}_0(\sqrt{(x_1 + y_1)^2 + k^2}).
$$
Here we can estimate
$$
   \mathbf{K}_0(x_1 + y_1)  \le  \left\{
   \begin{array}{ll}
     c\, (1 + |\log (x_1 + y_1)|), & 0 < x_1 + y_1 \le 1, \\
     c\, \ee^{-x_1 - y_1}, & x_1 + y_1 \ge 1.
   \end{array}\right.
$$
For $k \in \N$ we have $\sqrt{(x_1 + y_1)^2 + k^2} \ge 1$ for all $x_1,
y_1 \ge 0$ and thus
$$
   \mathbf{K}_0(\sqrt{(x_1 + y_1)^2 + k^2}) \le c\, \ee^{-\sqrt{(x_1 + y_1)^2 +
       k^2}}.
$$
It is now easy to see that $K \restriction (S'_+ \times S'_+)$ is square integrable:
indeed, it is easy to estimate
the contribution coming from the region $0 < x_1 + y_1 \le 1$ (note that only the case
$k=0$ comes with a logarithmic contribution).
As for the region $x_1 + y_1 \ge 1$, we may use the elementary estimate
\begin{align}
  \int_0^\infty\int_0^\infty \rom{e}^{-2\sqrt{(x_1 + y_1)^2 + k^2}} \d x_1 \d y_1
    & \le \int_0^k t \rom{e}^{-2k} dt + \int_k^\infty t \rom{e}^{-2t} \d t \nonumber\\
    & \le \frac{5}{4} k^2 \rom{e}^{-2k}, \qquad k \in \N. \nonumber
  \end{align}
The other cases can be treated in a similar fashion.

(5) In this step, we add in a (bounded) potential $W \ge 0$. Here the
Trotter product formula can be applied as in \cite[Prop.~1.3(b)]{V88} to
yield an inequality for the semigroups.
Pointwise information for the integral kernels is also obtained directly
as in the proof of Thm.~7 in \cite{DS76} via the Feynman-Kac formula which implies that
$$
 0 \le  \ee^{-t(H_{S'} + W)} (x,y) -  \ee^{-t(H_{S',\dec} + W)} (x,y)
  \le  \ee^{-t H_{S'}} (x,y) - \ee^{-t H_{S',\dec} } (x,y),
$$
for all $x, y \in S'$, and then the corresponding inequality holds for the resolvent
kernels, i.e.,
\begin{align}
  0 \le (H_{S'} + W + 1)^{-1}(x,y) - & (H_{S',\dec} + W + 1)^{-1}(x,y) \nonumber\\
   &  \le  (H_{S'} + 1)^{-1}(x,y) - (H_{S',\dec} + 1)^{-1}(x,y),
   \nonumber
\end{align}
which is in $\mathsf L_2(S'\times S')$ by step (4). We thus see that
$(H_{S'} + W + 1)^{-1} - (H_{S',\dec} + W + 1)^{-1}$ is Hilbert-Schmidt; furthermore,
there is a bound on the Hilbert-Schmidt norm of $(H_{S'} + W + 1)^{-1} -
(H_{S',\dec} + W + 1)^{-1}$ which is independent of $W \ge 0$. This proves Lemma~\ref{lem2.1}
for $r = 1$ and $W \ge 0$.

(6) If we consider $(-n,n) \times \S'$ instead of $\R \times \S'$, we may
again use path integral methods to show that
$$
  0 \le  \ee^{-t(L_{(-n,n)} + W)} (x,y) - \ee^{-t(L_{(-n,0)} \oplus L_{(0,n)} + W)} (x,y)
   \le  \ee^{-t H_S} (x,y) - \ee^{-t H_{S,\dec} } (x,y),
$$
so that $(L_{(-n,n)} + W + 1)^{-1} - (L_{(-n,0)} \oplus L_{(0,n)} + W +
1)^{-1}$ is Hilbert-Schmidt with a bound on the Hilbert-Schmidt norm that is
independent of $n$ and $W \ge 0$. This proves Lemma~\ref{lem2.2} for $r = 1$.
\subsection{Proof of Lemma~\ref{lem2.4}}
(1) As in Lemma~3.1 of \cite{vdBHV}, let $f \colon \R \to \R$ be
a bounded, monotonically increasing function of class $\mathsf C^2$ with
$|f'(x)| \le m$ and $|f''(x)| \le m$ for some constant $m$. We write
$H := L + W$ with arbitrary $W$ as in Lemma~\ref{lem2.4}.
Then
$$
    \ee^{-f} H \ee^{f} = H - 2 f' \partial_1 - f'' -
    |f'|^2.
$$
We first observe that for any $\eta > 0$
$$
   \lnorm{\nabla u}^2 \le \scapro{Hu}{u}
   \le \frac{\eta }{ 2 } \lnorm{H u}^2  + \frac{1 }{ 2\eta} \lnorm{u}^2,
    \qquad u \in D(L).
$$
Thus  the term $f' \partial_1$ is relatively bounded with
respect to $H$ with relative bound zero; the constants in
the estimate can be chosen to be independent of the potentials $W$
under consideration.
Write $H_f := \ee^{-f} H \ee^f$ with $D(H_f) = D(H)$.
As in \cite[Lemma 3.1]{vdBHV} there exists an $\eps_0 > 0$ (which is independent of
$W$) such that $H_{\eps f} - \lambda$ is invertible for all
$|\eps| \le \eps_0$ and all $\la \in [a,b]$  with a bound
$$
    \lnorm{(H_{\eps f} - \lambda)^{-1}} \le c_0, \qquad
    \lambda \in [a,b], \quad |\eps| \le \eps_0.
$$
Furthermore, $(H_{\eps f} - \lambda)^{-1} = \ee^{-\eps f} (H - \la)^{-1}
\ee^{\eps f}$.

(2) With $\xi_{2k}^+$ and $\xi_{2k}^-$ denoting the characteristic functions of the sets $(2k,\infty)\times\S'$
and $(-\infty,-2k)\times \S'$, respectively, we have  $1 - \chi_{2k} = \xi_{2k}^+ + \xi_{2k}^-$.
We then have
\begin{align}
\lnorm{\xi_{2k}^+ (H - \la)^{-1} \chi_k}
   & = \lnorm{\xi_{2k}^+ \ee^{\eps f} \ee^{-\eps f}(H -
     \la)^{-1}\ee^{\eps f}\ee^{-\eps f} \chi_k} \nonumber\\
   & \le  \lnorm{\xi_{2k}^+ \ee^{\eps f}}
        \lnorm{\ee^{-\eps f}(H - \la)^{-1}\ee^{\eps f}}
   \lnorm{\ee^{-\eps f} \chi_k}, \nonumber
\end{align}
with $\lnorm{\xi_{2k}^+ \ee^{\eps f}}$ denoting the norm of the operator
of multiplication by the function $\xi_{2k}^+ \ee^{\eps f}$.
We now specifically pick $f$ such that $f(x) = - x$ for $|x| \le 2k$ and we
let $\eps = \eps_0$. Then
$ \lnorm{\ee^{-\eps f}(H - \la)^{-1}\ee^{\eps f}} \le c_0$ by part (1), and
$ \lnorm{\ee^{-\eps f} \chi_k} \le \ee^{\eps_0 k}$, while
$ \lnorm{\xi_{2k}^+ \ee^{\eps f}} \le \ee^{-2k \eps}$. Dealing with
$\xi_{2k}^-(H - \la)^{-1} \chi_k$ in a similar fashion, we obtain that
\beq\label{6.2.2}
  \lnorm{(1 - \chi_{2k}) (H - \la)^{-1} \chi_k} \le c_1 \ee^{-\eps_0 k} .
\eeq
(3) We now turn to an eigenfunction $u$ of $L_{\R \setminus \{0\}} + W$
associated with an eigenvalue $\la \in [a,b]$. Let $\phi_k$ and $\psi_k = 1 - \phi_k$ be
as in {\red Section~\ref{sec_prelim}}. Then $\psi_k u \in D(H)$ with
$$
  (H - \la) (\psi_k u) = (L_{\R \setminus \{0\}} + W - \la) (\psi_k u)
   = -2 \psi_k' \partial_1 u - \psi_k'' u =: \Phi_k,
$$
whence, using $\supp(\Phi_k) \subset \{k/4 \le |x| \le k/2\}$ and $\lnorm{\Phi_k}
\le (c/k) \lnorm{u}$, we get
$$
  (1 - \chi_{k})u =  (1 - \chi_{k}) (H - \la)^{-1} \chi_{k/2} \Phi_k
$$
{\red and the desired estimate follows from \eqref{6.2.2}}.
\endproof
%
\subsection{Proof of Lemma~\ref{lem2.5}}
If the statement were not true we could find a sequence of
bounded, non-negative potentials $W_n$ such that $L + W_n$ has no spectrum in
the interval $(a_0,b_0)$ while $L_{\R \setminus \{0\}} + W_n$ has at least $n$ eigenvalues
in $[a,b]$ (counting multiplicities). Then, for each $n$, there exists an ONS of
eigenfunctions $u_{n,j}$, $j =1, \ldots, n$, associated with eigenvalues of
$L_{\R \setminus \{0\}} + W_n$ in $[a,b]$. With $\phi_k$ and $\psi_k = 1 - \phi_k$
as in Section~\ref{sec_main} we let $\mathcal{M}_{n,k}$ denote the subspace spanned by the functions
$\phi_{4k} u_{n,j}$, $j = 1, \ldots, n$. Let $C$ and $\gamma$ denote the constants in Lemma 2.4.
We then have the following:
\vskip1ex

{\bf Claim.} For $k = k_n := \lceil(\log n)/\gamma \rceil$ and $n$ large,
$\mathcal{M}_{n,k}$ has dimension $n$ and there exists a constant $M \ge 0$ such that
$$
    \lnorm{\nabla u}^2 \le M \lnorm{u}^2, \qquad  u \in \mathcal{M}_{n,k}.
$$
\vskip.5ex
By min-max, it follows from this claim that
$ \red{\tr \E_{(-\infty,M]}(L_{(-2 k_n, 2 k_n)}) \ge n}$
for $n$ large, in contradiction to Weyl's Law by which
$\red{\tr \E_{(-\infty,M]}(L_{(-2 k_n, 2 k_n)})}$ is bounded by $c \log n$.

Let us now prove the claim: any $u \in \mathcal{M}_{n,k}$ can be written as
$u = \phi_{4k} \sum_{j=1}^n \alpha_j u_{n,j}$ with suitable $\alpha_j \in \C$. Here we first note
that $v := \sum_j \alpha_j u_{n,j}$ has norm $ \norm{v} = (\sum_j |\alpha_j|^2)^{1/2}$
and satisfies $\norm{\nabla v}^2 \le b \norm{v}^2$ since the functions $u_{n,j}$,
$j = 1, \ldots, n$, form an ONS of eigenfunctions of $L + W_n$ with eigenvalues $\le b$.
We next observe that, by Lemma~\ref{lem2.4},
$$
   \norm{\psi_{4k} u_{n,j}} \le \norm{(1 - \chi_k) u_{n,j}} \le C \ee^{-\gamma k} \le C/n,
$$
for $j = 1, \ldots, n$, with $\chi_k$ denoting the characteristic function of the
set $(-k,k)\times \S'$.
Using the Schwarz inequality we get
$$
   \norm{(1 - \chi_k) v}
   \le \left({\textstyle \sum_j} |\alpha_j|^2\right)^{1/2}
   \cdot \left(\textstyle \sum_j \norm{(1 - \chi_k) u_{n,j}}^2 \right)^{1/2}
   \le \norm{v} (n C^2/n^2)^{1/2} = \frac C {\sqrt n} \norm{v},
$$
so that also $\norm{\psi_{4k}v} \le  \frac C {\sqrt n} \norm{v}$.
We then find for $n$ large,
$$
   \norm{u} = \norm{\phi_{4k} v} = \norm{v - \psi_{4k} v}
   \ge \norm{v} - \norm{\psi_{4k} v} \ge (1 - \frac C {\sqrt n}) \norm{v}
   \ge \frac 1 2 \norm{v}.
$$
In particular, the functions $\psi_{4k} u_{n,j}$, $j = 1, \ldots, n$ are
linearly independent for $n$ large. Furthermore,
 \begin{align}
  \norm{\nabla u} = \norm{\nabla (\phi_{4k}v)}
   \le \norm{\nabla\phi_{4k}}_\infty \norm{(1 - \chi_k) v} + \norm{\nabla v}
      \le \frac c {k n} + \sqrt b \norm{v}
\nonumber
 \end{align}
so that $\norm{\nabla u}^2 \le b_0 \norm{v}^2$ for $n$ large.
In view of the inequality $\norm{u} \ge  {\textstyle \frac 1 2} \norm{v}$, obtained above,
 we have therefore shown that, for $n$ large and $k = k_n$, $\norm{\nabla u}^2 \le 2 b_0 \norm{u}^2$ for
all $u \in \mathcal{M}_{n,k}$.
\endproof

\end{document}